%% file: main.tex
\documentclass{vldb}
\usepackage{graphicx}
\usepackage{balance}  
\pdfoutput=1

\usepackage{enumitem}
\usepackage{framed}
\usepackage{cprotect}
\usepackage{listings}
\usepackage{amstext}
\usepackage{pdfpages}
\usepackage{alltt}
\usepackage{epstopdf}
\usepackage{xspace,colortbl}
\usepackage[hyphens]{url}
\usepackage{amssymb}
\usepackage{fmtcount}
\usepackage{amsfonts}  
\usepackage{xspace}
\usepackage{amsmath}
\usepackage{multirow}
\usepackage{multicol}
\usepackage{parcolumns}
\usepackage[mathscr]{eucal}
\usepackage{colortbl}
\usepackage{lipsum}
\usepackage{dsfont}  
\usepackage{siunitx}
\usepackage{amsmath,amssymb}
\usepackage{algorithm}
\usepackage{algorithmicx}
\usepackage[noend]{algpseudocode}
\usepackage[labelfont=bf,font={small}]{caption}

\usepackage{bm}
\usepackage{cite}
\usepackage{csquotes}

\usepackage{adjustbox}

\definecolor{light-gray}{gray}{0.95}
\definecolor{mid-gray}{gray}{0.85}
\definecolor{green}{RGB}{0,176,80}
\definecolor{darkred}{rgb}{0.7,0.25,0.25}
\definecolor{darkgreen}{rgb}{0.15,0.55,0.15}
\definecolor{darkblue}{rgb}{0.1,0.1,0.5}
\definecolor{orange}{RGB}{237,125,49}
\definecolor{blue}{RGB}{68,114,196}
\definecolor{pop}{RGB}{0,21,245}
\definecolor{myblue}{RGB}{59,106,144}
\definecolor{realblue}{RGB}{0,0,255}

\usepackage{hyperref}       
\hypersetup{
    colorlinks=true,
    filecolor=magenta,
    linkcolor=darkred,
    citecolor=darkred,
    urlcolor=darkred,
}

\usepackage{tikz}
\usepackage{verbatim}
\usepackage{subcaption}
\newsavebox{\imagebox}
\usepackage{booktabs}
\usepackage{tabularx}

\usepackage{pifont}

\usepackage{bbm}

\newcommand{\ra}[1]{\renewcommand{\arraystretch}{#1}}
\newcommand{\secref}[1]{\S\ref{#1}}

\DeclareMathOperator*{\argmax}{arg\,max}

\newcommand{\sys}{\textsf{Naru}\xspace}
\newcommand{\pg}{\textsf{Postgres}\xspace}
\newcommand{\sqlserv}{\textsf{DBMS-1}\xspace}  
\newcommand{\kde}{\textsf{KDE}\xspace}

\newcommand{\dmv}{\textsf{DMV}\xspace}

\newcommand{\conviva}{\textsf{Conviva}\xspace}

\newtheorem{theorem}{Theorem}

\newcommand{\phat}{\ensuremath{\widehat{P}}\xspace}
\newcommand{\smallphat}{\ensuremath{\widehat{p}}\xspace}

\let\oldsim\sim
\renewcommand{\sim}{{\oldsim}}

\newcommand{\etal}{{\emph{et al.}\xspace}}

\newcommand{\revision}[1]{#1}
\newcommand{\removedfromrevision}[1]{}
\newcommand{\removedfromcamready}[1]{}

\newcommand{\titleph}{Deep Unsupervised Cardinality Estimation}
\vldbTitle{\titleph}
\vldbAuthors{Zongheng Yang, Eric Liang, Amog Kamsetty, Chenggang Wu, Yan Duan,
  Xi Chen, Pieter Abbeel, Joseph M. Hellerstein, Sanjay Krishnan, and Ion Stoica}
\vldbDOI{https://doi.org/10.14778/3368289.3368294}
\vldbVolume{13}
\vldbNumber{3}
\vldbYear{2019}

\definecolor{ReviewerDarkGray}{HTML}{37474F}
{\endlist}

\begin{document}



\title{\titleph}

\vspace{-.5in}
\numberofauthors{1}
\author{
  \alignauthor
  Zongheng Yang$^{\hspace{.1em}1}$, Eric Liang$^{\hspace{.1em}1}$, Amog Kamsetty$^{\hspace{.1em}1}$, Chenggang Wu$^{\hspace{.1em}1}$, Yan Duan$^{\hspace{.1em}3}$, \\Xi Chen$^{\hspace{.1em}1,3}$,
  Pieter Abbeel$^{\hspace{.1em}1,3}$, Joseph M. Hellerstein$^{\hspace{.1em}1}$, Sanjay Krishnan$^{\hspace{.1em}2}$, Ion Stoica$^{\hspace{.1em}1}$ \\
 \vspace{.4em}
\affaddr{$^1$UC Berkeley\hspace{0.5cm}$^2$University of
  Chicago\hspace{0.5cm}$^3$covariant.ai} \\
\vspace{4pt}
{\small
 $^1$\texttt{\{zongheng,ericliang,amogkamsetty,cgwu,pabbeel,hellerstein,istoica\}@berkeley.edu}\\
$^2$\texttt{skr@uchicago.\hspace{0.125em}edu} \hspace{.2em} $^3$\texttt{\{rocky,peter\}@covariant.ai}
}
}



\maketitle

\begin{abstract}
\input{abstract}

\end{abstract}

\input{intro}
\input{background}

\input{construction}
\input{querying}
\input{eval}
\input{related}

\input{conclusion}



\pagebreak
\bibliographystyle{abbrv}
\bibliography{refs}  

\balance


\end{document}

%% file: abstract.tex
\removedfromcamready{
\noindent Selectivity estimation has long been grounded in statistical tools for
density estimation.  To capture the rich multivariate distributions of
relational tables, we propose the use of a new type of high-capacity
statistical model: deep autoregressive models.  However, direct application of these models leads to a limited estimator that is prohibitively expensive to evaluate for range and wildcard predicates.      To make a truly usable
estimator, we develop a Monte Carlo integration scheme on top of autoregressive models that can efficiently handle
range queries with dozens of filters or more.
}

Cardinality estimation has long been grounded in statistical tools for density estimation.
To capture the rich multivariate distributions of relational tables, we propose the use of a new type of high-capacity statistical model: deep autoregressive models.
However, direct application of these models leads to a limited estimator that is prohibitively expensive to evaluate for range or wildcard predicates.
To produce a truly usable estimator, we develop a Monte Carlo integration scheme on top of autoregressive models that can efficiently handle range queries with dozens of dimensions or more.

Like classical synopses, our
estimator summarizes the data without supervision. Unlike previous solutions,
we approximate the joint data distribution without any independence
assumptions.  Evaluated on real-world datasets and compared against real
systems and dominant families of techniques, our
estimator achieves single-digit multiplicative error at tail, an up to $90\times$
accuracy improvement over the second best method, and is space- and
runtime-efficient.

%% file: intro.tex
\section{Introduction}

\removedfromcamready{
Estimating the selectivity of a SQL predicate
is a core primitive
in query optimization~\cite{systemr}, approximate query processing~\cite{blinkdb}, and performance profiling~\cite{akdere2012learning}.  The main task is to estimate the fraction of a relation that
satisfies a predicate, without actual execution.
Despite its importance, there is wide agreement that the problem is
still unsolved~\cite{leis2015good,leis2018query,reopt}.
Open-source and commercial DBMSes
routinely produce up to $10^4{-}10^8\times$ estimation errors on queries over a large number of attributes~\cite{leis2015good}.
}

Cardinality estimation is a core primitive in query optimization~\cite{systemr}.
One of its main tasks is to accurately estimate the \emph{selectivity} of a SQL predicate---the
fraction of a relation selected by the predicate---without actual execution.
Despite its importance, there is wide agreement that the problem is
still unsolved~\cite{leis2015good,leis2018query,reopt}.
Open-source and commercial DBMSes
routinely produce up to $10^4{-}10^8\times$ estimation errors on queries over a large number of attributes~\cite{leis2015good}.

The fundamental difficulty of selectivity estimation comes from condensing information
about data into summaries~\cite{space_synopsis}.  The predominant approach in database systems today
is to collect single-column summaries (e.g., histograms and sketches), and to
combine these coarse-grained models assuming column independence.  This
represents one end of the spectrum, where the summaries are fast to construct and cheap to
store, but compounding errors occur due to the coarse information and over-simplifying
independence assumptions.
On the other end of the spectrum, when given the {\em joint data distribution} of
a relation
(the frequency of each unique tuple normalized by the relation's cardinality),
perfect selectivity ``estimates'' can be read off or computed via integration over the distribution. However,
the {\em joint} is intractable to compute or store for all but the tiniest datasets.
Thus, traditional selectivity estimators face the hard
tradeoff between the amount of information captured
and the cost to construct, store, and query the summary.

\begin{figure}[t]
\centering
\includegraphics[width=0.8\columnwidth]{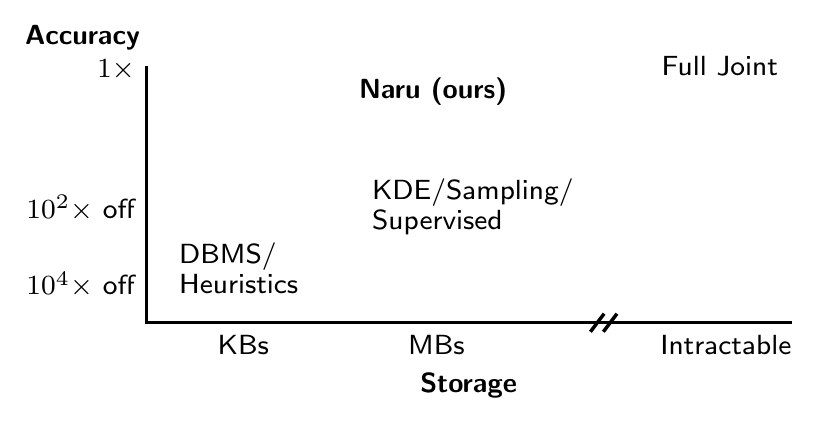}
\caption{Approximating the joint data distribution in full, \sys enjoys high
  estimation accuracy and space efficiency.\label{fig:teaser}}
\end{figure}

An accurate and compact {\em joint approximation} would allow better design points in this tradeoff space (Figure~\ref{fig:teaser}).
Recent advances in deep unsupervised learning have offered promising tools in this regard. 
While it was previously thought intractable to approximate the data
distribution of a relation in its full
form~\cite{getoor2001selectivity,deshpande2001independence},
{\em deep autoregressive models}, a type of density estimator,
have
succeeded in modeling high-dimensional data such as images, text, and audio~\cite{pixelcnn,pixelcnnpp,wavenet, transformer}.
However, these models only estimate {\em point densities}---in query processing
terms, they only handle equality predicates (e.g., ``what is the fraction of tuples with \textsf{price} equal to \$100?'').
Full-featured selectivity estimation requires handling not only equality but
also range predicates (e.g., ``what fraction of tuples have \textsf{price} less
than \$100 and \textsf{weight} greater than \SI{10}lbs?'').
Naive estimation of the range density by integrating over the query region
requires summing up an enormous number of points.
In an 11-dimensional table we consider, a challenging range query has $10^{10}$ points in the
query region, which would take more than 1,000 hours to sum over by a naive enumeration scheme.
A full-featured selectivity estimator, therefore, requires new techniques beyond
the state of the art.

In this paper, we show that selectivity estimation can be performed with high accuracy by using deep autoregressive models.
We first show how relational data---including both numeric and categorical attributes---can be mapped onto these models for effective selectivity estimation of equality predicates.
We then introduce a new Monte Carlo integration technique called
\emph{progressive sampling}, which efficiently estimates range queries even at high dimensionality.
By leveraging the availability of conditional probability distributions provided
by the model,
progressive sampling steers the sampler into regions of high probability
density, and then corrects for the induced bias by using importance
weighting.
This technique extends the state of the art in density estimation, with particular applicability to our problem of general-purpose selectivity estimation.
Our scheme is effective: a thousand samples suffice to accurately estimate the
aforementioned $10^{10}$-point query.
To realize these ideas, we design and implement \sys (\underline{N}eur\underline{a}l \underline{R}elation \underline{U}nderstanding), 
a selectivity estimator that approximates the joint data
distribution in its full form, without any column independence assumptions. 
Approximating the joint in full not only provides superior accuracy, but also
frees us
from specifying what combinations of columns to build synopses on.
We further propose optimizations to efficiently handle wildcard predicates, and to encode and decode real-world relational data
(e.g., supporting various datatypes, small and large domain sizes).
Combining our integration
scheme with these practical strategies results in a highly accurate, compact,
and functionality-rich selectivity estimator based on deep autoregressive models.
Just like classical synopses, \sys summarizes a relation in an unsupervised fashion.
The model is trained via statistically grounded principles (maximum likelihood) where no supervised signals or query feedback are required.
While \emph{query-driven} estimators are optimized with respect to a set of training queries
(i.e., ``how much error does the estimator incur on these queries?''),
\sys is optimized with respect to the underlying data distribution (i.e., ``how divergent is the estimator from the data?'').
Being data-driven, \sys supports a much larger set of queries and is automatically robust to query distributional shifts.  Our evaluation
compares \sys to the state-of-the-art unsupervised and supervised techniques, showing \sys to be
the only estimator to achieve worst-case \textit{single-digit multiplicative errors} for
challenging high-dimensional queries.
\removedfromcamready{Our full joint approximation is orthogonal to
query-driven approaches: an unsupervised autoregressive model can always take
advantage of additional signals such as query feedback by further fine-tuning.  On the other hand,
}

In summary, we make the following contributions:
\begin{enumerate}
    \item We show deep autoregressive models can be used for selectivity
      estimation (\secref{sec:background}, \secref{sec:likelihood-models}), and propose optimizations
      to make them suitable for relational data (\secref{sec:construction}).
    \item To handle challenging range queries, we develop
      \textit{progressive sampling}, a Monte Carlo integration algorithm that
      efficiently estimates range densities even with large query regions (\secref{sec:prog-sampling}).
      We augment it with a novel
      optimization, \emph{wildcard-skipping} (\secref{sec:wildcard-skipping}),
      to handle wildcard predicates.
       We also propose information-theoretic column orderings
      (\secref{sec:column-ordering}) to reduce estimation variance.

    \item We extensively evaluate on real datasets against 8
      baselines across 5 different families (heuristics, real DBMSes, sampling, statistical methods, deep supervised regression).
      Our estimator \sys achieves up to
      orders-of-magnitude better accuracy with space usage $\sim 1\%$
      of data size and $\sim 5{-}10\si{\ms}$ of estimation latency (\secref{sec:eval}).


\end{enumerate}


%% file: background.tex
\section{Problem Formulation}
\label{sec:background}
Consider a relation $T$ with
attribute domains $\{A_1,\dots,A_n\}$.
Selectivity estimation seeks to estimate the fraction of tuples in $T$ that satisfy
a particular predicate, $\theta: A_1\times\cdots\times A_n \to \{0,1\}$. We define the selectivity to be
$\textsf{sel}(\theta) := {|\{\bm{x} \in T: \theta(\bm{x}) = 1 \}|} / {|T|}$.

The {\em joint data distribution} of the relation, defined to be
\[
P(a_1, \dots, a_n) := f(a_1,\dots, a_n) / |T| 
\]
 is closely related to the selectivity, where $f(a_1,\dots, a_n)$ is the number of occurrences of tuple $(a_1, \dots,
a_n)$ in $T$.  It forms a valid probability distribution since integrating it over the attribute domains yields a value of $1$.
Thus, exact selectivity calculation is equivalent to integration over the joint:
\[
\textsf{sel}(\theta) = \sum_{a_1 \in A_1}\dots  \sum_{a_n \in A_n} \theta(a_1,\dots,a_n) \cdot
P(a_1,\dots,a_n).
\]
In this work, we consider finite relation $T$ and hence its empirical
domains $A_i$ are finite. Therefore summation is used in the integration calculation
above.



\subsection{Approximating the Joint via Factorization}
\label{sec:joint-approx}
Given the joint, exact selectivity ``estimates'' can be calculated by integration.
However, the number of entries in the joint---and thus the maximum number of points
needed to be summed over in the integration---is $|P| = \prod_{i = 1}^n |A_i|$,
a size that grows exponentially in the number of attributes.
Real-world tables with a
dozen or so columns can easily have a theoretic joint size of $10^{20}$ and upwards
(\secref{sec:eval}).
\revision{In practice, it is possible to bound this number by $|T|$, the number of
  tuples in the relation, by not storing any entry with zero occurrence.}
Algorithmically, to scale construction, storage, and integration to high-dimensional tables, joint approximation techniques seek to \textit{factorize} \cite{grimmett2001probability} the joint into some lower-dimensional representation, $\phat \approx P$.

Classical 1D histograms~\cite{systemr} use the simplest factorization,
$\phat(A_1, \cdots, A_n) \approx \prod_{i = 1}^n \phat(A_i)$, where independence between attributes is assumed.
The $\phat(A_i)$'s are materialized as histograms that are cheap to construct and store.
Selectivity estimation reduces to calculating per-column selectivities and combining by multiplication,
\[
\textsf{sel}(\theta) \approx \left(  \sum_{a_1 \in A_1} \theta_1(a_1) \phat(a_1)\right) \times \cdots \times\left( \sum_{a_n \in A_n} \theta_n(a_n) \phat(a_n) \right)
\]
where each $\theta_i$ is predicate $\theta$ projected to each attribute
(assuming here $\theta$ is a conjunction of single-attribute filters). 

Richer factorizations are possible and are generally more accurate.  For instance, Probabilistic Relational
Models~\cite{getoor2001selectivity,getoor2001learning} from the early 2000s leverage the conditional independence assumptions of Bayesian Networks
(e.g., joint factored
into smaller distributions, \(\{ \phat(A_1 | A_2, A_3),$ $\phat(A_2), \phat(A_3)\}\)).
Dependency-Based Histograms~\cite{deshpande2001independence} use decomposable
interaction models and rely on partial independence between columns (e.g.,
$\phat(A_1, A_2, A_3) \approx \phat(A_1) \phat(A_2, A_3)$).  Both
methods are marked improvements over 1D histograms since they capture more than
single-column interactions. However, the tradeoff between richer factorizations and costs to store
or integrate is still unresolved.
Obtaining selectivities becomes drastically harder due to the integration
now crossing multiple attribute domains. Most importantly, the approximated joint's
precision is compromised since some forms of independence are still assumed.
\removedfromrevision{The above parametric factors are materialized either as probability tables or as
histograms, thus the cost of construction and storage grows large for a large number of attributes.}

In this paper, we consider the richest possible factorization of the joint, using the product rule:
\[
\phat(A_1, \cdots, A_n) = \phat(A_1) \phat(A_2 | A_1) \cdots \phat(A_n | A_1,\dots,A_{n-1})
\]
Unlike the previous proposals, the product rule factorization is an {\em exact}
relationship to represent a distribution.  It makes no
independence assumptions and captures all complex interactions between attributes.
Key to this goal is that the factors, $\{\phat(A_i | A_1,\dots,A_{i-1})\}$, need not be materialized; instead, they are calculated
on-demand by a neural network, a high-capacity universal function approximator~\cite{esl}.

\subsection{Problem Statement}
\label{subsec:problem}
We estimate the selectivities of queries of the following form.
A query is a conjunction of single-column boolean predicates, over arbitrary
subsets of columns.  A predicate contains an attribute, an operator,
and a literal, and is read as $A_i \in R_i$ (attribute $i$ takes on
values in valid region $R_i$).
Our formulation includes the usual $=, \neq, <, \leq, >, \geq$ predicates, the rectangular
containment $A_i \in [l_i, r_i]$, or even \textsf{IN} clauses.  For ease of
exposition, we use \emph{range} to denote the valid region $R_i$ or, for the
whole query, the composite valid region $R_1 \times \cdots \times R_n$.
We assume the \emph{domain} of each column, $A_i$, is finite:  since a real dataset
is finite, we can take the empirically present values of a column as its finite domain.



We make a few remarks.  First,
\removedfromrevision{the supported queries
are wide: the usual $=, \neq, <, \leq, >, \geq$ operators, the rectangular
containment $A_i \in [l_i, r_i]$, or even the \textsf{IN} operator are
considered ranges under our formulation.}
disjunctions of such predicates are supported via the inclusion-exclusion principle.
Second, our formulation follows a large amount of existing work on this
topic~\cite{deshpande2001independence,getoor2001selectivity,selectivity-kde,poosala1997selectivity,quicksel}
and, in some cases, offers more capabilities. Certain prior work
requires each predicate be a rectangle~\cite{kipf2018learned,selectivity-kde} or
columns be real-valued~\cite{korn1999range,selectivity-kde};
our ``region'' formulation supports complex predicates and does
not make these assumptions.
Lastly,
the relation under
estimation can either be a base table or a join result. \\

\iftrue
\fi

\section{Deep Autoregressive Models}
\label{sec:likelihood-models}
\subsection{Overview}
\removedfromcamready{
\sys uses a deep likelihood model to approximate the joint distribution.
We overview two categories of such models and discuss the subset
suitable for a selectivity estimator.\\
}
\sys uses a deep autoregressive model to approximate the joint distribution.
We overview the statistical features they offer and how those relate to selectivity estimation.\\

\removedfromcamready{
{\noindent \bf Access to point density $\phat(\bm{x})$.}  All deep likelihood
models can produce point density estimates $\phat(\bm{x})$ after training on a set
of n-dimensional tuples $T = \{\bm{x}_1, \cdots\}$ with the unsupervised maximum likelihood objective.
Many network architectures have been proposed,
ranging from
latent variable models~\cite{vae}, 
normalizing flows~\cite{realnvp,glow}, to autoregressive models such as masked multi-layer
perceptrons (MLP)~\cite{uria13_deep_tract_densit_estim,made} and
the Transformer~\cite{transformer}. \\
}
{\noindent \bf Access to point density $\phat(\bm{x})$.}  Deep autoregressive
models produce point density estimates $\phat(\bm{x})$ after training on a set
of n-dimensional tuples $T = \{\bm{x}_1, \dots\}$ with the unsupervised maximum likelihood objective.
Many network architectures have been proposed in recent years,
such as masked multi-layer perceptrons
(e.g., MADE~\cite{made}, ResMADE~\cite{resmade}) or masked self-attention
networks (e.g., Transformer~\cite{transformer}). \\

\removedfromcamready{
{\noindent \bf Access to autoregressive densities $\{\phat(x_i | \bm{x}_{< i})\}$.}  A subclass of
models, termed {\em autoregressive models}, provides access to the conditional densities present in the product rule:
}

{\noindent \bf Access to conditional densities $\{\phat(x_i | \bm{x}_{< i})\}$.}
Additionally, autoregressive models also provide access to all conditional densities present in the product rule:
\begin{equation*}
\begin{aligned}
   \phat(\bm{x}) & = \phat(x_1, x_2, \cdots, x_n) \\
  & = \phat(x_1) \phat(x_2 | x_1) \cdots \phat(x_n | x_1,\dots,x_{n-1})
\end{aligned}
\end{equation*}
Namely, given input tuple
$\bm{x}=(x_1, \cdots, x_n)$, one can obtain from the model the $n$ conditional density estimates,
$\{\phat(x_i | \bm{x}_{< i})\}$.
The model can be architected to use any ordering(s) of the attributes (e.g.,
$(x_1, x_2, x_3)$ or $(x_2, x_1, x_3)$). In our exposition we assume
the left-to-right schema order (\secref{sec:column-ordering} discusses heuristically picking a good ordering). \\

\removedfromcamready{At first glance, the additional conditionals may not seem to benefit our selectivity
estimation task,
but there are two important reasons this family of models is
the right choice.}
\sys chooses autoregressive models for selectivity estimation for two
important reasons.
First, autoregressive models have shown superior modeling precision
in learning images~\cite{pixelcnn,pixelcnnpp}, audio~\cite{wavenet}, and
text~\cite{transformer}.  All these domains involve correlated, high-dimensional data akin to a
relational table.
Second, as we will show in \secref{sec:prog-sampling}, access to conditional densities
is critical in efficiently supporting range queries.



\subsection{Autoregressive Models for Relational Data}
\label{sec:implement-autoreg}
\begin{figure}[tp]
\centering
\includegraphics[width=.95\columnwidth]{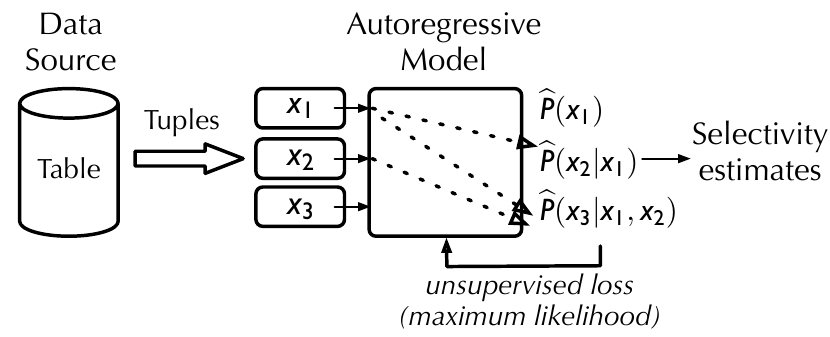}
\vspace{.1in}
\caption{{{Overview of the estimator framework}.
    \sys is trained by reading data tuples and does not require supervised training
    queries or query feedback, just like classical synopses.
  }}
\label{fig:arch}
\end{figure}

\sys allows any autoregressive model $\mathcal{M}$ to be plugged in.
In general, such model has the following functional form:
  \begin{equation}
  \mathcal{M}(\bm{x}) \mapsto \left[ \phat(X_1), \phat(X_2 | x_1), \cdots, \phat(X_n | x_1,\dots,x_{n-1}) \right]
  \label{eq:autoreg}
\end{equation}
Namely, one tuple goes in, a list of
conditional density distributions comes out, each being a {\em distribution} of the $i$th attribute
conditioned on previous attributes. (The \emph{scalars} required to compute the point
density, $\{\phat(x_i | \bm{x}_{< i})\}$, are read from these conditional distributions.)
How can a neural net $\mathcal{M}$ attain the autoregressive property, e.g., that 
 $\phat(X_3 | x_1, x_2)$ only depends on, or ``sees'', the information from the
first two attribute values $(x_1, x_2)$ but not anything else?

 \emph{Information masking} is a common technique used to implement
autoregressive models\cite{made,transformer,pixelcnn}; here we illustrate the
idea by constructing an example architecture for relational data.
Suppose we assign
each column $i$ its own compact neural net, whose input is the aggregated
information about {\em previous} column values $\bm{x}_{<i}$. Its role is to
use this context information to output
a distribution over its own domain,  $\phat(X_i | \bm{x}_{<i})$.  Consider a ${\tt travel\_checkins}$ table 
with columns ${\tt{city}}, {\tt year}, {\tt stars}$.
Assume the model is given the input tuple, $\langle {\tt{Portland}},\allowbreak {\tt 2017},\allowbreak {\tt 10} \rangle$.
First, column-specific encoders $E_{\tt{col}}()$ transform each attribute value
into a numeric vector suitable for neural net consumption,
$[E_{\tt{city}}({\tt{Portland}}),\allowbreak E_{\tt year}({\tt{2017}}), \allowbreak E_{\tt stars}({\tt{10}})]$.
Then, appropriately aggregated inputs are fed to the per-column neural nets
$\mathcal{M}_{\tt col}$:
\begin{gather*}
  \bm{0} \rightarrow \mathcal{M}_{\tt{city}} \\ 
  E_{\tt{city}}({\tt{Portland}}) \rightarrow \mathcal{M}_{\tt{year}} \\
 \oplus \left(  E_{\tt{city}}({\tt{Portland}}), E_{\tt year}({\tt{2017}}) \right) \rightarrow \mathcal{M}_{\tt{stars}} 
\end{gather*}
where $\oplus$ is the operator that aggregates information from
several encoded attributes.  In practice, this aggregator can be
vector concatenation, a set-invariant {\em pooling}
operator (e.g., elementwise sum or max), or even self-attention~\cite{transformer}.

Notice that the first output, from $\mathcal{M}_{\tt{city}}$, does not depend on
any attribute values (its input $\bm{0}$ is arbitrarily chosen).  The second output depends only on the attribute
value from ${\tt city}$, and the third depends only on both ${\tt city}$ and ${\tt
  year}$.  Therefore, the three outputs can be interpreted as
\[
  \left[ \phat({\tt{city}}), \phat({\tt{year}} | {\tt{city}}), \phat({\tt{stars}} | {\tt{city}}, {\tt{year}})\right]
\]
Thus, autoregressiveness is achieved via such input masking.

Training these model outputs to be as close as possible to the true conditional
densities 
is done via maximum likelihood estimation.
Specifically, the {\em cross entropy}~\cite{esl} between the data distribution $P$ and
the model estimate $\phat$ is calculated over all tuples in relation $T$ and used as the loss:
\begin{equation}
  \mathcal{H}(P, \phat) = - \sum_{{\bm x} \in T} P({\bm x}) \log \phat({\bm x}) \\
= - \frac{1}{|T|} \sum_{{\bm x} \in T} \log \phat({\bm x})
\label{eq:loss}
\end{equation}
It can be fed into a standard gradient descent optimizer~\cite{adam}. Lastly,
the Kullback-Leibler divergence, $\mathcal{H}(P, \phat) - \mathcal{H}(P)$, is the
\emph{entropy gap} (in bits-per-tuple) incurred by the model.
A lower gap indicates a higher-quality density estimator; thus, it serves as a monitoring metric
during and after training.

%% file: construction.tex
\section{Estimator Construction}
\label{sec:construction}

We now discuss practical issues in constructing \sys.

\subsection{Workflow}
\label{subsec:overview}


Figure~\ref{fig:arch} outlines the workflow of building a \sys estimator.
After specifying a table $T$ to build an estimator on,
batches of random tuples from $T$ are read to train \sys. 
In practice, a snapshot of the table
can be saved to external storage so normal DBMS activities are not
affected. Neural network training can be performed either close to the
data (at periods of low activity) or offloaded to a remote
process. 

For a batch of tuples, \sys encodes each attribute value using column-specific strategies (\secref{sec:encoding-decoding}).  The encoded batch then gets fed into the model to
perform a gradient update step.    Our evaluation
(\secref{subsec:training}) empirically observed that one pass over
data is sufficient to achieve a high degree of accuracy (e.g., outperforming
real DBMSes by $10{-}20\times$), and more passes are beneficial until model convergence.


Appends and updates may cause statistical staleness.
\sys can be fine-tuned on the updated relation to correct for this, as we show in \secref{subsec:refresh}.
\revision{
Further, if new data comes in per-day partitions, then each partition can train its own \sys model.
Efficient incremental model update is an important topic worthy of detailed
study, which we defer to future work.} \\

{\noindent \bf Joins.}  The estimator does not distinguish between the type of
table it is built on.  To build an estimator on a joined relation, either the entire joined relation can be pre-computed and materialized,
or multi-way join operators~\cite{triejoin,viglas2003maximizing} and samplers~\cite{leis2017cardinality,chaudhuri1999random} can be used to
produce batches of tuples on-the-fly.  Given access to tuples from the joined
result, no
changes are needed to the estimator framework.  Once trained, the
estimator supports queries that filter any column in the joined
relation.
This treatment follows prior work~\cite{neo,quicksel,kiefer2017estimating} and is conceptually clean.

\subsection{Encoding and Decoding Strategies}
\label{sec:encoding-decoding}
\sys models a relation as a high-dimensional discrete distribution.  The key
challenge is to {\em encode} each column into a form suitable for neural
network consumption, while preserving the column semantics.  Further, each
column's output distribution $\phat(X_i | {\bm x}_{<i})$ (a vector of scores) must be efficiently
{\em decoded} regardless of its datatype or domain size.




For each column \sys first obtains its domain $A_i$ either
from user annotation or by scanning. All values in the column are then
dictionary-encoded into integer IDs in range $[0, |A_i|)$.  For
instance, the dictionary can be ${\tt Portland} \mapsto
0$, ${\tt SF} \mapsto 1$, etc.  For
a column with a natural order, e.g., numerics or strings, the domain is
sorted so that the dictionary order follows the column order.
\revision{Overall, this pre-processing step is a lossless transformation (i.e.,
  a bijection).}
\removedfromrevision{
 A special placeholder $\bot$ can be inserted into the domain so that a
 previously built estimator can function on new data.
}

Next, column-specific encoders $E_{\tt col}()$ encode these IDs into vectors.  The ML community has proposed many such strategies
before; we make sensible choices by keeping in mind a few
characteristics specific to relational datasets: \\

{\noindent \bf Encoding small-domain columns: one-hot.}
For such a column $E_{\tt col}()$ is set to
{\em one-hot encoding} (i.e., indicator variables).
For instance, if there are a total of 4 cities, then
the encoding of ${\tt SF}$ is $E_{\tt city}(1) = [0, 1, 0, 0]$, a
4-dimensional vector.
The small-domain threshold is configurable and
set to 64 by default.  This encoding takes $O(|A_i|)$ space per value. \\

{\noindent \bf Encoding large-domain columns: embedding.}
For a larger domain, the one-hot vector wastes space and computation budget.
\sys uses embedding encoding in this case.
In this scheme---a preprocessing step in virtually all natural language processing tasks---a learnable embedding matrix of type $\mathbb{R}^{|A_i| \times h}$ is
randomly initialized, and $E_{\tt col}()$ is simply row lookup into this
matrix. For instance, $E_{\tt year}(4)
\mapsto$ row 4 of embedding matrix, an $h$-dimensional vector.   The
embedding matrix gets updated during gradient descent as part of the model
weights.
Per value this takes $O(h)$ space (\sys defaults $h$ to 64).
This encoding
is ideal for domains with a meaningful semantic distance (e.g., cities are
similar in geo-location, popularity, relation to its nation) since each
dimension in the embedding vector can learn to represent each such similarity. \\

{\noindent \bf Decoding small-domain columns.} Suppose domain $A_i$ is small.
In this easy case, the network allocates an {\em output layer} to compute a {\em distribution} $\phat(X_i |
\bm{x}_{<i})$, which is a $|A_i|$-dimensional vector of probabilities used for
selectivity estimation.
We use a fully connected layer,
$\textsf{FC}(F, |A_i|)$, where $F$ is the hidden unit size.
For example, for a city column with three values in its domain, the output distribution may be $\left[\textsf{SF}=0.2; \textsf{Portland}=0.5; \textsf{Waikiki}=0.3\right]$.
During optimization, the training loss seeks to minimize the divergence of this output from the data distribution.  \\

{\noindent \bf Decoding large-domain columns: embedding reuse.}  If the domain is large, however,
using a fully connected output layer $\textsf{FC}(F, |A_i|)$ would be inefficient in both
space and compute.
Indeed, an ${\tt id}$ column in a dataset we
tested on has a large domain size of $|A_i| = 10^4$, inflating the
output layer beyond typical scales.

\sys solves this problem
by an optimization that we call
``embedding reuse''.  In essence, we replace the potentially large output layer
$\textsf{FC}(F, |A_i|)$ with a much smaller version, $\textsf{FC}(F, h)$ (recall
that $h$ is the typically small embedding dimensions; defaults to 64).  This
immediately yields a saving ratio of $|A_i|/h$.
The goal of decoding is to take in
inputs $\bm{x}_{<i}$ and output $|A_i|$ probability scores over the domain.
With the shrunk-down output layer,
inputs $\bm{x}_{<i}$ would pass through the net
arriving at an $h$-dimensional feature vector, $H \subseteq \mathbb{R}^{1 \times
  h}$.
We then
calculate $H E_i^T$, where $E_i \subseteq \mathbb{R}^{|A_i|\times h}$ is the
{\em already-allocated} embedding matrix for column $i$,  
obtaining a vector $\mathbb{R}^{1 \times |A_i|}$ that can be interpreted as
the desired scores after normalization.
We have thus decoded the output while cutting down the cost of compute and storage.
This scheme has proved effective in other
large-domain tasks
~\cite{gpt2}. 










\subsection{Model Choice}
\label{subsec:model-choice}
As discussed, any autoregressive model can be plugged in, taking
advantage of \sys's encoding/decoding optimizations as well as querying
capabilities (\secref{sec:querying}). %
We experiment with three representative architectures:
(A) Masked Autoencoder (MADE)~\cite{made}, a standard multi-layer perceptron with
information masking to ensure autoregressiveness;
(B) ResMADE~\cite{resmade}, a simple extension to MADE where residual
connections are introduced to improve learning efficiency; and
(C) Transformer~\cite{transformer}, a class of self-attentional models driving
recent state-of-the-art advances in natural language
processing~\cite{bert,xlnet}.
Table~\ref{table:ar-choice} compares the tradeoffs of these building blocks.  We found that, under similar parameter
count, more advanced architectures (B, C) achieve better entropy gaps; however, the
smaller entropy gaps do not automatically translate into better selectivity
estimates and the computational cost can be significantly higher (for C).

\removedfromrevision{
We provide several implementations to choose from:
(A) the architecture outlined in \secref{sec:implement-autoreg}
and (B) a masked autoencoder~\cite{made} (a basic multi-layer perceptron with
similar information-limiting to ensure autoregressiveness).
On a 15-column dataset we evaluated on, when controlling for the same parameter count, architecture A achieves better quality than B (8\% better entropy gap; \secref{sec:bits-gap}).
The same trend holds on another dataset.  However, architecture
B's current implementation is more optimized.  \sys therefore defaults to
architecture B over better learning quality and faster convergence.  We defer the
study of all possible model choices to future work.
}%


%% file: querying.tex
\section{Querying the Estimator}
\label{sec:querying}


Once an autoregressive model is trained, it can be queried to compute selectivity
estimates.  Assume a query $\textsf{sel}(\theta) = P(X_1 \in R_1, \dots, X_n \in R_n)$
asking for the selectivity of the conjunction,
where each range $R_i$ can be a point (equality predicate), an interval (range
predicate), or any subset of the domain (\textsf{IN}).  The calculation of this density
is fundamentally summing up the probability masses distributed in the
cross-product region, $R = R_1 \times \cdots \times R_n$.

We first discuss the straightforward support for equality predicates, then move
on to how \sys solves the more challenging problem of range predicates. \\

\noindent {\bf Equality Predicates}.  When values are specified for {\em all}
columns, estimating conjunctions
of these equality predicates is straightforward.
Such a point query has the form $P(X_1 =
x_1, \dots, X_n = x_n)$
and requires only a single
forward pass on the point, $(x_1, \dots, x_n)$, to obtain the sequence of
conditionals, $[\phat(X_1 = x_1), \phat(X_2 = x_2 | X_1 = x_1), \dots, \phat(X_n
= x_n | X_1 = x_1, \dots, X_{n-1}=x_{n-1}) ]$, which are then multiplied.\\

\noindent {\bf Range Predicates}. It is impractical to assume a workload that
only issues point queries.  With the presence of any range predicate, or
when some columns are not filtered, the number of points that must be evaluated through the model becomes larger than $1$.
(In fact, it easily grows to an astronomically large number for the majority of workloads we considered.)
We discuss two ways in which \sys carries out this operation.
\textit{Enumeration} exactly sums up the densities when the
queried region $R$ is sufficiently small:
\removedfromrevision{all discrete points in this region are
enumerated and fed into the model, in a batching fashion, whose corresponding
point densities are then summed up:}
 \[
\textsf{sel}(X_1 \in R_1, \dots, X_n \in R_n) \approx \sum_{x_1 \in R_1}\dots\sum_{x_n \in R_n}\phat(x_1, \dots, x_n).
 \]

 When the region $R$ is deemed too
big---almost always the case in the datasets and workloads we considered---we instead
use a novel approximate technique termed \textit{progressive sampling}
(described next), an
unbiased estimator that works surprisingly well on the relational
datasets we considered.  

\revision{Lastly, queries with out-of-domain literals can be handled via simple rewrite.
For example, suppose $\textsf{year}$'s domain is $\{ 2017, 2019 \}$.  A range
query with an out-of-domain literal, say ``$\textsf{year} < 2018$'', can be
rewritten as ``$\textsf{year} \leq 2017$'' with equivalent semantics.
For equality predicates with out-of-domain literals, \sys simply returns a
cardinality of 0.  Hereafter we consider in-domain literals and valid regions.}


\subsection{Range Queries via Progressive Sampling}
\label{sec:prog-sampling}

\begin{figure}[tp]
\centering
\includegraphics[width=1\columnwidth]{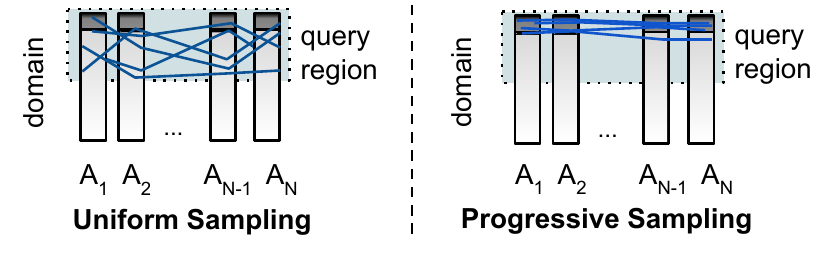}
\caption{{{The intuition of progressive sampling.}  Uniform samples taken from
the query region have a low probability of hitting the high-mass sub-region of the
query region, increasing the variance of Monte Carlo estimates. Progressive
sampling avoids this by sampling from the estimated data distribution instead, which naturally concentrates samples in the high-mass sub-region.}}
\label{fig:uniformvsps}
\end{figure}

The queried region $R = R_1 \times \cdots \times R_n$ in the worst case contains
$O(\prod_{\substack{i}} D_i)$ points, where $D_i=|A_i|$ is the size of each
attribute domain.  Clearly, computing the likelihood for an exponential number
of points is prohibitively expensive for data/queries with even moderate
dimensions.  \sys proposes an approximate integration scheme to address this challenge.  \\

\noindent {\bf First attempt} (Figure~\ref{fig:uniformvsps}, left).  The simplest way to approximate the sum is via
uniform sampling.  First, sample $\bm{x}^{(i)}$ uniformly at random from $R$.
Then, query the model to compute $\smallphat_i = \phat(\bm{x}^{(i)})$.  By naive
Monte Carlo, for $S$ samples we have
$\frac{|R|}{S} \sum_{i = 1}^S \smallphat_i$ as an
unbiased estimator to the desired density.
Intuitively, this scheme is randomly throwing points into target region $R$ to
probe its average density. \\

To understand the failure mode of uniform sampling,
consider a relation $T$ with $n$ correlated columns, with each column distribution
skewed so that $99\percent$ of the probability mass is contained in the top
$1\percent$ of its domain (Figure \ref{fig:uniformvsps}). Take a query with range predicates selecting the
top $50\percent$ of each domain. It is easy to see that uniformly sampling from the query region will take in
expectation $1/(0.01 / 0.5)^n = 1/0.02^n$ samples to hit the high-mass region we
are integrating over. Thus, the number of samples needed for an accurate
estimate increases exponentially in $n$. Consequently, we find that
this sampler collapses catastrophically in the real-world datasets that we
consider.
It has the worst errors among all baselines in our evaluation.
\\



\noindent {\bf Progressive sampling} (Figure~\ref{fig:uniformvsps}, right).
Instead of uniformly throwing points into the region, we could be more selective
in the points we choose---precisely leveraging the power of the trained autoregressive model.
Intuitively, a sample of the first dimension $x_1^{(i)}$ would allow
us to \textit{``zoom in'' into the more meaningful region of the second
  dimension}.  This more meaningful region is exactly described by the second
conditional output from the autoregressive model, $\phat(X_2 | x_1^{(i)})$, a distribution over the
second domain given the first dimension sample.  We can obtain a sample of the
second dimension, $x_2^{(i)}$, from this space instead of from
$\text{Unif}(R_2)$. This sampling process continues for all columns.
To summarize, progressive sampling consults the autoregressive model to steer the
sampler into the high-mass part of the query region, and finally compensating for the induced bias with importance weighting.
\\

{\noindent \em Example.} We show the sampling procedure for a 3-filter query.
Drawing the $i$-th sample for query $P(X_1 \in R_1, X_2 \in R_2, X_3 \in R_3)$:
\begin{enumerate}
    \item Forward $\bm{0}$ to get $\phat(X_1)$.
      Compute and store \(\phat(X_1 \in R_1)\) by summing.
      Then draw $x_1^{(i)} \sim \phat(X_1 | X_1
      \in R_1)$.
    \item Forward $x_1^{(i)}$ to get $\phat(X_2|x_1^{(i)})$. Compute and store $\phat(X_2 \in R_2 | x_1^{(i)})$. Draw $x_2^{(i)} \sim \phat(X_2 | X_2 \in R_2, x_1^{(i)})$.   
    \item Forward $(x_1^{(i)}, x_2^{(i)})$ to get $\phat(X_3|x_1^{(i)},
      x_2^{(i)})$.
      Compute and store $\phat(X_3 \in R_3 | x_1^{(i)}, x_2^{(i)})$.  
\end{enumerate}
The summation and sampling steps are fast since they are only over single-column distributions. This is in contrast to integrating or summing over all columns at once, which has an exponential number of points. The product of the three stored intermediates,
\begin{equation}\label{eq:three-col-progressive}
\phat(X_1 \in R_1) \cdot \phat(X_2 \in R_2 | x_1^{(i)}) \cdot \phat(X_3 \in R_3 | x_1^{(i)}, x_2^{(i)})
\end{equation}
is an unbiased estimate for the desired density.
By construction, the sampled point satisfies the query ($x_1^{(i)}$ is drawn from range $R_1$, $x_2^{(i)}$ from $R_2$, and so forth).  It remains to show that this sampler is approximating the correct sum:

\begin{algorithm}[t]
\caption{Progressive Sampling: estimate the density of query region $R_1 \times \cdots \times R_n$ using $S$ samples.}\label{alg:progressive-sampling}
\begin{algorithmic}[1]
\Function{ProgressiveSampling}{$S; R_1, \dots, R_n$}
\State $\phat = 0$
\For{$i= 1 \text{ to } S$}\Comment{Batched in practice}
\State $\phat = \phat + \textproc{Draw}(R_1, \dots, R_n)$
\EndFor
\State \textbf{return} $\phat / S$
\EndFunction

\Statex

\Function{Draw}{$R_1, \dots, R_n$}\Comment{Draw one tuple}
    \State $\smallphat = 1$
    \State $\bm{s} = \bm{0}_n$ \Comment{The tuple to fill in}
    \For{$i= 1 \text{ to } n$}
    \State Forward pass through model: $\mathcal{M}(\bm{s})$
    \State  $\phat(X_i | \bm{s}_{<i}) = $ the $i$-th model output  \Comment{Eq.~\ref{eq:autoreg}}
    \State Zero-out probabilities in slots $[0, D_i) \setminus R_i$
    \State Re-normalize, obtaining $\phat(X_i | X_i \in R_i, \bm{s}_{<i})$
    \State $\smallphat = \smallphat \times \phat(X_i \in R_i | \bm{s}_{<i})$

    \State Sample $s_i \sim \phat(X_i | X_i \in R_i, \bm{s}_{<i})$
    \State $\bm{s}[i] = s_i$
    \EndFor
    \State \textbf{return} $\smallphat$\Comment{Density of the sampled tuple $s$}
\EndFunction
\end{algorithmic}
\end{algorithm}

\begin{theorem}
\label{unbiased}
Progressive Sampling estimates are unbiased.
\end{theorem}

The proof only uses basic probability rules and is deferred to our online technical report.
Algorithm~\ref{alg:progressive-sampling} shows the pseudocode for the general
$n$-filter case.
For a column that does not have an explicit filter, it can in theory be treated as having
a wildcard filter, i.e., $R_i = [0, D_i)$.
We describe our more efficient treatment,
\emph{wildcard-skipping}, in \secref{sec:wildcard-skipping}.
Our evaluation shows that the sampler can cover both low and high density regions, and handles challenging
range queries for large numbers of columns and joint spaces.

\revision{
  Progressive sampling bears connections to sampling algorithms in graphical
  models.
  Notice that the autoregressive factorization corresponds to a complex
  graphical model where each node $i$ has all nodes with
  indices $<i$ as its parents.  In this interpretation, progressive sampling extends
  the \emph{forward sampling with likelihood weighting} algorithm\cite{pgm_textbook} to allow
  variables taking on \emph{ranges of values} (the former, in its default form, allows
  equality predicates only).
}

\removedfromrevision{This sampler bears some resemblance to
sequential importance sampling~\cite{doucet2001introduction} and may be amenable
to applicable analysis.   Progressive sampling has a sequential dependency between columns for drawing a single sample, but multiple samples can be drawn independently, which lends itself well for parallelization with modern hardware.} 

\removedfromrevision{
Lastly, we note that progressive sampling not only can operate on the model's
approximated joint, but Algorithm~\ref{alg:progressive-sampling} also works on
an oracle joint distribution (e.g., obtained by data scanning at each step).  We take advantage of
this observation in \secref{subsec:microbenchmarks} to run the sampler against
an oracle distribution. }


\revision{
\subsection{Reducing Variance: Wildcard-Skipping}
\label{sec:wildcard-skipping}


\sys introduces \emph{wildcard-skipping}, a simple optimization to efficiently
handle wildcard predicates. 
Instead of sampling through the full domain
of each wildcard column in a query, $X_i \in \ast$, we could restrict it to
a special token, $X_i = \textsf{MASK}_i$.
Intuitively, $\textsf{MASK}_i$
signifies column $i$'s \emph{absence} and essentially marginalizes it. 
In our experiments, wildcard-skipping can reduce the variance of worst-case errors
by several orders of magnitude (\secref{eval:variance-reduction}).

During training, 
we perturb each tuple so that the training data contains \textsf{MASK} tokens.
We uniformly sample a subset of columns to \emph{mask out}---their original
values in the tuple are discarded and replaced with corresponding
$\textsf{MASK}_{\text{col}}$.
For an $n$-column tuple, each column has a probability of $w/n$ to be masked out, where $w\,\sim\,\text{Unif}[0,n)$.
The output target for the cross-entropy loss still uses the original values.



\subsection{Reducing Variance: Column Ordering}
\label{sec:column-ordering}
\sys models adopt a single ordering of columns during construction (\secref{sec:construction}). However, different orderings may have different sampling efficiency. For instance, having ${\tt city}$ as the first column and setting it to ${\tt Waikiki}$ focuses on records only relevant to that city, a data region supposedly much narrower
than that from having ${\tt year}$ as the first column.

Empirically, we find that these heuristic orders work well:
\begin{enumerate}
  \item {\sf MutInfo}: successively pick column $X_i$ that maximizes the \emph{mutual information}~\cite{cover2012elements} between all columns chosen so far and itself, $\argmax_i I(X_{\text{chosen}}; X_i)$.
  \item {\sf PMutInfo}: a variant of the above that maximizes the \emph{pairwise} mutual information, $\argmax_i I(X_{\text{last}}; X_i)$.
\end{enumerate}


Intuitively, maximizing $I(X_{\text{chosen}}; X_i)$ corresponds to finding the
next column with \emph{the most information already contained in the chosen
  columns}. For both schemes, we find that picking the column with the maximum marginal entropy, $\argmax_i H(X_i)$, as the first works well.
Interestingly, on our datasets, the ${\sf Natural}$ ordering
(left-to-right order in table schema) is also effective. We hypothesize this is
due to human bias in placing important or ``key''-like columns earlier that
highly reduce the uncertainty of other columns.

Lastly, we note that \emph{order-agnostic training} has been proposed in the ML
literature~\cite{made,uria13_deep_tract_densit_estim,xlnet}. The idea is to train the same model on more
than one order, and at inference time invoke a (presumably seen) order most
suitable for the query. This is a possible future optimization for \sys, though
in the preliminary experiments we did not find the performance benefits on top
of our optimizations significant.


}


%% file: eval.tex
\section{Evaluation}\label{sec:eval}

We answer the following questions in our evaluation:
\begin{enumerate}
    \item How does \sys compare to state-of-the-art selectivity estimators in accuracy (\secref{subsec:accuracy})?  Is it robust (\secref{subsec:ood})?
    \item How long does it take to train a \sys model to achieve a useful level of accuracy (\secref{subsec:training})?
    \item \sys requires multiple inference passes to produce a selectivity estimate. How does this compare with the latency of other approaches (\secref{subsec:latency})?
    \item How do wildcard-skipping and column orderings affect accuracy and
      variance (\secref{eval:variance-reduction})?  How does accuracy change
      with model choices and sizes (\secref{subsec:space})?
\end{enumerate}
Lastly, a series of
    microbenchmarks are run
    to understand \sys's limits (\secref{subsec:microbenchmarks}).

\subsection{Experimental Setup}
We compare \sys against predominant families of selectivity estimation techniques,
including estimators in real databases, heuristics, non-parametric density estimators, and supervised learning approaches (Table~\ref{table:baselines}). To ensure a fair comparison between
estimators, we restrict each estimator to a fixed \textit{storage
  budget} (Table~\ref{table:datasets}). For example, for the \conviva-A dataset,
\sys's model must be less than 3MB in size, and the same restriction is held for
all estimators for that dataset when applicable.
\subsubsection{Datasets}

We use real-world datasets with challenging
characteristics (Table~\ref{table:datasets}). The number of rows ranges from 10K to 11.6M, the
number of columns ranges from 11 to 100, and the size of the joint space ranges
from $10^{15}$ to $10^{190}$: 

\textbf{\dmv~\cite{dmv-dataset}.} Real-world dataset consisting of vehicle
registration information in New York.  We use the following 11 columns with
widely differing data types and domain sizes
(the numbers in parentheses):
{\sf  record\_type} (4), {\sf reg\_class} (75), {\sf state} (89),
{\sf county} (63), {\sf body\_type} (59), {\sf fuel\_type} (9), {\sf
  valid\_date} (2101), {\sf color} (225), {\sf sco\_ind} (2), {\sf sus\_ind}
(2), {\sf rev\_ind} (2).  Our snapshot contains 11,591,877 tuples.  The exact joint distribution has a size of $3.4 \times 10^{15}$.


\textbf{\conviva-A.} Enterprise dataset containing anonymized user activity logs from a video analytics company. The table corresponds to 3 days
    of activities.  The 15 columns contain a mix of small-domain categoricals
    (e.g., error flags, connection types) as well as large-domain numerical
    quantities (e.g., various bandwidth numbers in kbps).  Although the domains
    have a range (2--1.9K) similar to \dmv, there are many more numerical
    columns with larger domains,
    resulting in a much larger joint distribution ($10^{23}$).

\textbf{\conviva-B.} A small dataset of 10K rows and 100 columns also from Conviva, with a joint space of over $10^{190}$. Though this dataset is trivial in size, this enables the use of an emulated, perfect-accuracy model for running detailed robustness studies (\secref{subsec:microbenchmarks}).

\subsubsection{Estimators}
We next discuss the baselines listed in Table~\ref{table:baselines}.

\textbf{Real databases.}
\textsf{Postgres} and {\sqlserv} represent the performance a practitioner can
hope to obtain from a real DBMS.  Both rely on classical assumptions
and 1D histograms, while the latter additionally contains cross-column
correlation statistics.  Every column has a histogram and associated statistics
built.
{\sf Postgres} is tuned to use a maximum amount of per-column bins (10,000). For
{\sqlserv}, one invocation of
stats creation with all columns
specified only builds a histogram on the first column; we therefore invoke
stats creation several times so that all columns are covered.



\textbf{Independence assumption.} {\sf Indep} scans each column to obtain perfect per-column selectivities and combines them by multiplication. This measures the inaccuracy solely attributed to the independence assumption.

\begin{table}[tp]\centering \small%
\ra{1.3}
\caption{{List of datasets used in evaluation.} ``Dom.'' refers to per-column domain size.  ``Joint'' is number of entries in the exact joint distribution (equal to the product of all domain sizes).  ``Budget'' is the storage budget we allocated to all evaluated estimators, when applicable, relative to the in-memory size of the corresponding original tables.\label{table:datasets}}
\begin{tabular}{@{} l l l l l l @{}} \toprule
{\textsc{Dataset}} & {\textsc{Rows}} & {\sc Cols} & {\sc Dom. } & \textsc{Joint} & {\textsc{Budget}}  \\ \midrule
\dmv & 11.6M & 11 & 2--2K & $10^{15}$ & $1.3\%$ (13MB) \\
\conviva-A & 4.1M & 15 & 2--1.9K & $10^{23}$ & $0.7\%$ (3MB) \\ 
\conviva-B & 10K & 100 & 2--10K & $10^{190}$ & N/A \\ 
\bottomrule
\end{tabular}
\end{table}

\begin{table}[tp]\centering \small%
\ra{1.3}
\caption{\small{List of estimators used in evaluation.}\label{table:baselines}}
\begin{tabular}{@{} l l p{4.6cm} @{}} \toprule
{\textsc{Type}} & {\textsc{Estimator}}  & {\textsc{Description}}  \\ \midrule
\removedfromrevision{Histogram & \textsf{Hist} & N-dimensional histogram. \\}
Heuristic & {\sf Indep} & A baseline that multiplies \textit{perfect} per-column selectivities.\\
Real System & \textsf{Postgres} & 1D stats and histograms via independence/uniformity assumptions.\\
Real System & \sqlserv & Commercial DBMS: 1D stats plus inter-column unique value counts.\\
  Sampling &
\textsf{Sample}
& Keeps $p\%$ of all tuples in memory.  Estimates a new query by evaluating on those samples. \\
\revision{MHIST} & \revision{{\sf MHIST}} & The \textsf{MaxDiff(V,A)} histogram~\cite{poosala_improved}.\\
\revision{Graphical} & \revision{\textsf{BayesNet}} & Bayes net (Chow-Liu tree~\cite{chowliu}).\\
KDE& \textsf{KDE} & Kernel density estimation~\cite{selectivity-kde,kiefer2017estimating}.\\ 
Supervised   & \textsf{MSCN} & Supervised deep regression net~\cite{kipf2018learned}.\\
Deep AR & \sys  & (Ours) Deep autoregressive models. \\ \bottomrule
\end{tabular}
\end{table}

\revision{
\textbf{Multi-dimensional histogram.}
We compare to MHIST, an N-dimensional histogram.
We use \textit{MaxDiff} as our partition constraint, \textit{Value} (V) as the sort parameter, and \textit{Area} (A) as the source parameter~\cite{poosala_improved}.
We use the MHIST-2 algorithm~\cite{poosala1997selectivity} and the \textit{uniform spread} assumption~\cite{poosala_improved} to approximate the value set within each partition.
According to~\cite{poosala1997selectivity}, the resulting
MaxDiff(V,A) histogram offers the most accurate and robust performance compared
to other state-of-the-art histogram variants.

\textbf{Bayesian Network.}
We use a Chow-Liu tree~\cite{chowliu} as the Bayesian Network, since empirically it gave the
best results for the allowed space. Since the size of conditional probability
tables scales with the cube of column cardinalities, we use equal-frequency
discretization (to 100 bins per column) to bound the space consumption and
inference cost.
Lastly, we apply the same progressive sampler to allow this estimator to support
range queries (it does not support range queries out of the box);
this ensures a fair comparison between the use of deep autoregressive models and this approach.}

\textbf{Kernel density estimators \& Sampling.}
In the non-parametric sampling regime, we evaluate a uniform sampler and a
state-of-the-art KDE-based selectivity estimator~\cite{selectivity-kde,kiefer2017estimating}.
\textsf{Sample}
keeps a set of $p\%$ of tuples
uniformly at random from the original table. In accordance with our memory
budget for each dataset, $p$ is set to $1.3\%$ for \dmv and $0.7\%$ for
\conviva-A. \textsf{KDE}~\cite{selectivity-kde} attempts to learn the underlying
data distribution by averaging Gaussian kernels centered around random sample
points. The number of sample points is chosen in accordance with our memory
budget: 150K samples for \dmv and 28K samples for \conviva-A. The bandwidth for
\textsf{KDE} is computed via Scott's rule \cite{scottrule}. The bandwidth for
\textsf{KDE-superv} is initialized in the same way, but is further optimized
through query feedback from 10K training queries. We modify the source code released by the authors~\cite{kde-code} in order to run it with more than ten columns.

\textbf{Supervised learning.}
We compare to a recently proposed supervised deep net-based estimator termed
\textit{multi-set convolutional network}~\cite{kipf2018learned}, or
\textsf{MSCN}. We apply the source code from the authors~\cite{github-kipf} to
our datasets.  As it is a \textit{supervised} method, we generate 100K training
queries from the same distribution the test queries are drawn, ensuring their
``representativeness''. The net stores a materialized sample of the data. Every
query is run on the sample to get a bitmap of qualifying tuples---this is used
as an input additional to query features. We try three variants of the model, all with the same hyperparameters and all trained to convergence:
\textsf {MSCN-base} uses the same setup reported
originally~\cite{kipf2018learned} (1K samples, 100K training queries) and consumes 3MB. 
We found that \textsf{MSCN}'s performance is highly dependent on the samples, so we include
a variant with $10\times$ more samples ({\sf MSCN-10K}: 10K samples, 100K train
queries), consuming 13MB (satisfying \dmv's budget only).
We also run \textsf{MSCN-0} that stores no samples and uses query features only.

\textbf{Deep unsupervised learning (ours).}
We train one \sys model for each dataset.  All models
are trained with the unsupervised maximum likelihood objective.  Unless
stated otherwise, we employ wildcard-skipping and the natural column ordering.
Sizes are reported without any compression of network weights: 
\begin{itemize}
\item \dmv: masked autoencoder (MADE), 5 hidden layers
  (512, 256, 512, 128, 1024 units), consuming 12.7MB. 

\item \conviva-A: MADE, 4 hidden layers with 128 units each, consuming 2.5MB. The embedding reuse optimization with $h=64$ is used (\secref{sec:encoding-decoding}).
  \removedfromrevision{
\item \conviva-B:
  Emulated ``oracle model''. This dataset is for microbenchmarks only.
  }
\end{itemize}
For timing experiments,
we train and run the learning methods ({\sf KDE}, {\sf MSCN}, \sys) on a
V100 GPU. Other estimators are run on an 8-core node
and vectorized when applicable.

\subsubsection{Workloads}
\label{sec:eval-workloads}
\begin{figure}[t]
\centering
\includegraphics[width=.9\columnwidth]{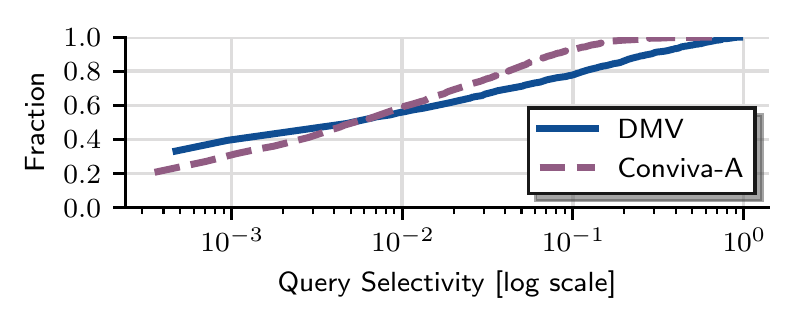}
\caption{\small {Distribution of query selectivity} (\secref{sec:eval-workloads}).\label{fig:sel-dist}}
\end{figure}
 {\noindent \bf Query distribution} (Figure~\ref{fig:sel-dist}).  We generate multidimensional queries
 containing both range and equality predicates.  The goal is to test each
 estimator on a wide spectrum of target selectivities: we group them as high (${>}2\%$),
 medium ($0.5\%$--$2\%$), and low ($\leq 0.5\%$).  Intuitively, all solutions should perform reasonably well for high-selectivity queries, because dense regions require only coarse-grained modeling capacity.  As the query selectivity drops, the estimation task becomes harder, since low-density regions require each estimator to model details in each hypercube.  True selectivities are obtained by executing the queries on {\sf Postgres}.
\begin{table*}[t!]\centering \small%
\ra{1.05}
\caption{{{Estimation errors on \dmv}.
    Errors are grouped by true selectivities and shown in percentiles computed from 2,000 queries.
}  \label{table:standalone-combined}}
\begin{tabular}{@{} l l l l l l l l l l l l l l l l @{}} \toprule
\textsc{Estimator} && \multicolumn{4}{c}{\sc High ($(2\%, 100\%]$)}  & & \multicolumn{4}{c}{\sc Medium ($(0.5\%, 2\%]$)} & & \multicolumn{4}{c}{\sc Low ($\leq 0.5\%$)} \\
  && Median  & 95th & 99th & Max  && Median  & 95th & 99th & Max && Median  & 95th & 99th & Max\\ \midrule


    
{\sf Indep} &&  1.12 & 1.55 & 46.1 & 2566
    &&  1.25 & 46.6 & 1051 &  $8 \cdot 10^4$
    &&  1.35 & 225 & 2231 & $2\cdot10^4$ \\


    
  {\sf Postgres}
                   &&  1.12 & 1.55 & 46.3 & 2608
    &&  1.25 & 45.5 & 1070 & $8 \cdot 10^4$
    &&  1.36 & 227 & 2287 & $2 \cdot 10^4$\\

    
{\sqlserv} &&  1.45 & 3.36 & 5.80 & 12.6
    && 2.72 & 6.38 & 9.29 & 10.1
    &&  5.28 & 83.0 & 417 & 917  \\

    
{\sf Sample}
&&  \textbf{1.00 }& \textbf{1.02} & \textbf{1.03} & \textbf{1.05}
    &&  {1.02} & {1.05} & \textbf{1.07} & \textbf{1.10}
    &&   1.12 & 43.2 & 98.1 & 377 \\



    
  {\sf KDE} &&   10.9 & 1502
                                                                       & $1 \cdot 10^4$ & $2 \cdot 10^5$
                                                                                                                           && 48.0 & $2 \cdot 10^4$ 
             & $9 \cdot 10^4$ & $2 \cdot 10^5$
    &&  38.0 & 3191 & $2 \cdot 10^4$ & $5 \cdot 10^4$  \\

    
{\sf KDE-superv} &&  1.40 & 3.81 & 4.91 & 13.3
    &&  1.53 & 4.36 & 8.12 & 16.9
    &&  1.95 & 30.0 & 98.0 & 175 \\

{\sf MSCN-base}
&& 1.17 & 1.42 & 1.47 & 1.58
    && 1.14 & 1.65 & 2.53 & 3.96
    && 2.95 & 32.5 & 85.6 & 921 \\
    
{\sf MSCN-0}
&& 16.8 & 92.4 & 195 & 285
    && 8.89 & 80.4 & 344 & 471
    && 4.79 & 67.1 & 169 & 6145\\
    
    
{\sf MSCN-10K}
&& 1.04 & 1.10 & 1.12 & 1.16
    && 1.04 & 1.12 & 1.19 & 1.23
    && 1.51 & 14.2 & 33.7 & 264 \\
    
    
\revision{{\sf MHIST}}
&& \revision{1.70} & \revision{2.65} & \revision{4.11} & \revision{9.20}
    && \revision{1.65} & \revision{6.00} & \revision{15.1} & \revision{21.1}
    && \revision{2.81} & \revision{70.3} & \revision{352} & \revision{5532}  \\




    \revision{{\sf BayesNet}} && \revision{1.01} & \revision{1.07} & \revision{1.12} & \revision{1.44}
&& \revision{1.05} & \revision{1.18} & \revision{1.26} & \revision{1.40}
&& \revision{3.41} & \revision{16.1} & \revision{79} & \revision{561}  \\

    \midrule



\sys-1000 && 1.01 & 1.03 & 1.05 & 1.28
&& \textbf{1.01} & 1.06 & 1.15 & 1.27
&& \textbf{1.03} & 1.44 & 2.51 & \textbf{8.00} \\
\sys-2000 && 1.01 & 1.03 & 1.04 & 1.16
&& \textbf{1.01} & \textbf{1.04} & 1.09 & 1.38
&& \textbf{1.03} & \textbf{1.41} & \textbf{2.18} & \textbf{8.00} \\

\bottomrule
\end{tabular}
\vspace{0.25em}

\end{table*}

\begin{table*}[t!]\centering \small%
\ra{1.05}
\caption{{{Estimation errors on \conviva-A}.
    Errors grouped by true selectivities and shown in percentiles computed from 2,000 queries.
}  \label{table:standalone-combined-conviva}}
\begin{tabular}{@{} l l l l l l l l l l l l l l l l @{}} \toprule
\textsc{Estimator} && \multicolumn{4}{c}{\sc High ($(2\%, 100\%]$)}  & & \multicolumn{4}{c}{\sc Medium ($(0.5\%, 2\%]$)} & & \multicolumn{4}{c}{\sc Low ($\leq 0.5\%$)} \\
  && Median  & 95th & 99th & Max  && Median  & 95th & 99th & Max && Median  & 95th & 99th & Max\\ \midrule


    
{\sqlserv} && 1.75 & 5.25 & 7.77 & 9.12
    && 3.93 & 13.6 & 19.9 & 31.3
    && 8.63 & 176 & 636 & 4737 \\
    

    
{\sf Sample} && \textbf{1.02} & \textbf{1.06} & \textbf{1.09} & \textbf{1.11}
    && \textbf{1.04} & \textbf{1.14} & \textbf{1.18} & \textbf{1.23}
    && 1.18 & 49.3 & 218 & 696\\
{\sf KDE} && 105 & $2 \cdot 10^5$ & $5 \cdot 10^5$ & $8 \cdot 10^5$
    && 347 & $6 \cdot 10^4$ & $8 \cdot 10^4$ & $8 \cdot 10^4$
    && 224 & $1 \cdot 10^4$ & $2 \cdot 10^4$ & $2 \cdot 10^4$ \\
{\sf KDE-superv} && 1.99 & 7.97 & 14.5 & 33.6
    && 2.04 & 8.44 & 17.0 & 49.7
    && 2.76 & 74.5 & 251 & 462 \\
{\sf MSCN-base} && 1.14 & 1.27 & 1.36 & 1.48
    && 1.15 & 1.55 & 2.26 & 57.7
    && 2.05 & 20.3 & 84.1 & 370 \\
    
\revision{{\sf MHIST}} && \revision{1.54} & \revision{5.24} & \revision{11.2} & \revision{$1 \cdot 10^4$}
    && \revision{2.22} & \revision{10.5} & \revision{30.3} & \revision{$3 \cdot 10^4$}
    && \revision{6.71} & \revision{792} & \revision{4170} & \revision{$8 \cdot 10^4$} \\





 \revision{{\sf BayesNet}}  && \revision{1.15} & \revision{1.75} & \revision{2.05} & \revision{2.70}
&& \revision{1.31} & \revision{2.70} & \revision{4.95} & \revision{47.6}
&& \revision{1.67} & \revision{14.8} & \revision{78.0} & \revision{998}  \\

    \midrule

\sys-1000&& \textbf{1.02} & 1.11 & 1.18 & 1.37
&& 1.05 & 1.17 & 1.27 & 1.40
&& 1.10 & 1.71 & 3.01 & 185  \\

\sys-2000&& \textbf{1.02} & 1.10 & 1.16 & 1.28
&& 1.05 & 1.17 & 1.27 & 1.38
&& \textbf{1.09} & 1.66 & 3.00 & 58.0  \\

\sys-4000&& \textbf{1.02} & 1.10 & 1.17 & 1.21
&& \textbf{1.04} & 1.15 & 1.27 & 1.36
&& \textbf{1.09} & \textbf{1.57} & \textbf{2.50} & \textbf{4.00}  \\

\bottomrule
\end{tabular}
\vspace{0.25em}

\vspace{-.2in}
\end{table*}

The query generator is inspired by prior
work~\cite{kipf2018learned}. Instead of designating a few fixed columns to
filter on, we consider the more challenging scenario where filters are
randomly placed.
First, we draw the number of (non-wildcard) filters $5 \leq f \leq 11$ uniformly at random.
We always include at least five filters to avoid queries with very high
selectivity, on which all estimators perform similarly well.
Next, $f$ distinct columns are drawn to place the filters.
For columns with domain size $\geq 10$, the filter operator is sampled uniformly
from $\{=, \leq, \geq\}$; for columns with small domains,
the equality
operator is picked---the intention is to avoid placing a range predicate on
categoricals, which often have a low domain size. 
The filter literals are then chosen from a random tuple sampled uniformly from
the table, i.e., they follow the data distribution. For example, a
  valid 5-filter query on \dmv is ``$({\sf fuel\_type}=\text{GAS}) \land ({\sf rev\_ind} = \text{N}) \land ({\sf sco\_ind} = \text{N}) \land ({\sf valid\_date} \geq \text{2018-03-23}) \land ({\sf color} = \text{BK})$''.
Overall, the queries span a wide range
of selectivities (Figure~\ref{fig:sel-dist}). \\


{\noindent \bf Accuracy metric.}  We report accuracy by the multiplicative error~\cite{kipf2018learned,leis2015good,leis2018query} (also termed ``Q-error''),
the factor by which an estimate differs from the actual cardinality:
\[
\text{Error} := \max(estimate, actual) /  \min(estimate, actual)
\]
We lower bound the estimated and actual cardinalities at 1
to guard against division by zero.  In line with prior work~\cite{deshpande2001independence,leis2015good}, we found that the multiplicative error is much more informative than the \textit{relative} error, as the latter does not fairly penalize small cardinality estimates (which are frequently the case for high-dimensional queries). Lastly we report the errors in quantiles, with a particular focus at the tail. Our results show that all estimators can achieve low median (or mean) errors but with greatly varying performance at the tail, indicating that mean/median metrics do not accurately reflect the hard cases of the estimation task.



\subsection{Estimation Accuracy}
\label{subsec:accuracy}


In summary, Tables \ref{table:standalone-combined} and
\ref{table:standalone-combined-conviva} show that not only does \sys match or
exceed the best estimator across the board, it excels in the \textit{extreme tail of query difficulty}---that is, worst-case errors on low-selectivity queries.
For these types of queries, \sys achieves orders of magnitude better accuracy
than classical approaches, and up to $90\times$ better tail behavior than query-driven (supervised) methods.

The same \sys model is used to estimate all queries on a dataset, showing the robustness of the model learned. We now discuss these macrobenchmarks in more detail.

\begin{figure*}[htp]
\centering
     \centering
     \begin{subfigure}[b]{0.49\textwidth}
         \centering
\includegraphics[width=\columnwidth]{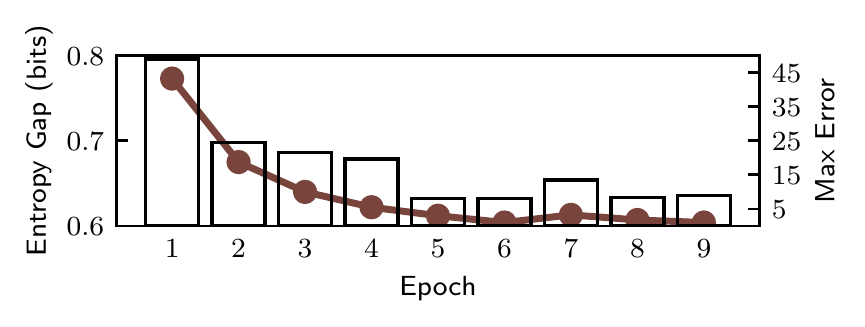}
\vspace{-.25in}
         \caption{\small{\dmv}}
     \end{subfigure}
     \hfill
     \begin{subfigure}[b]{0.49\textwidth}
         \centering
\includegraphics[width=\columnwidth]{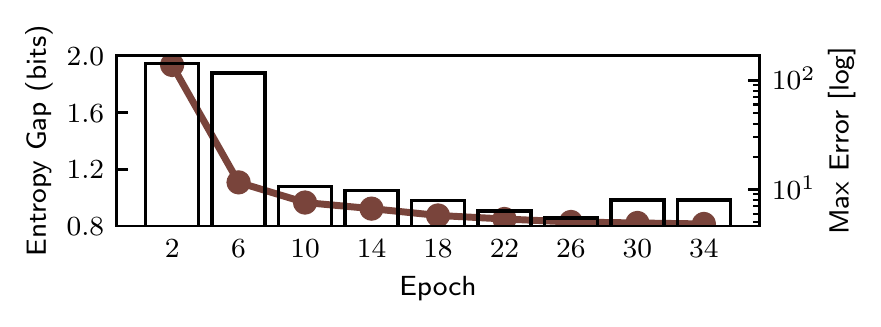}
\vspace{-.25in}
         \caption{\small{\conviva-A}}
     \end{subfigure}
\vspace{.1in}
\caption{{{Training time vs. quality} (\secref{subsec:training}). Dotted lines show divergence from data; bars show max estimation errors.}}
\label{fig:dmv-train-time-vs-divergence}
\end{figure*}

\begin{figure*}[htp]
     \centering
     \begin{subfigure}[b]{0.49\textwidth}
         \centering
\includegraphics[width=\columnwidth]{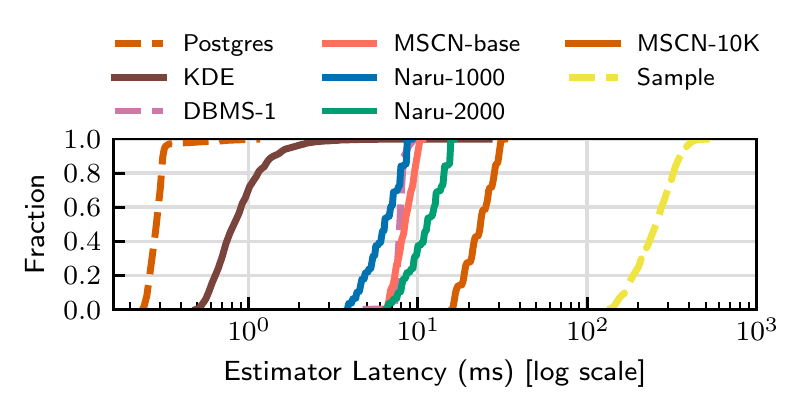}
\vspace{-.25in}
         \caption{\small{\dmv}}
     \end{subfigure}
      \hfill
     \begin{subfigure}[b]{0.49\textwidth}
         \centering
\includegraphics[width=\columnwidth]{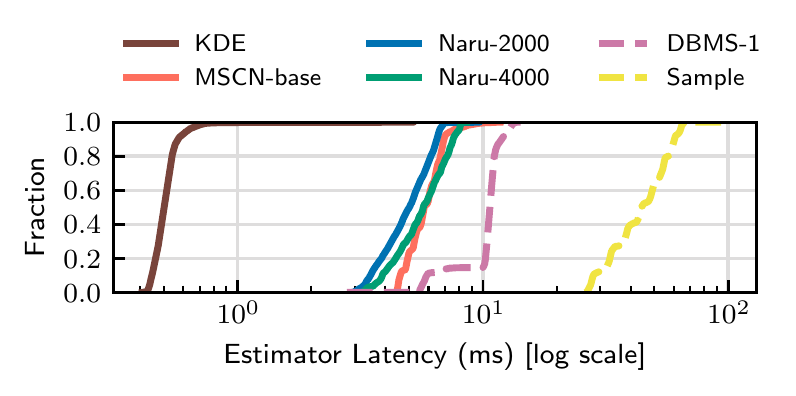}
\vspace{-.25in}
\caption{\small{\conviva-A}}
     \end{subfigure}
\vspace{.1in}
\caption{{{Estimator latency} (\secref{subsec:latency}).  Learning
    methods are run on GPU; other estimators are run on CPU (dashed
    lines). \removedfromrevision{Legend order follows line order; best viewed in color.}}}
\label{fig:latency}
\end{figure*}

\subsubsection{Results on \dmv}

\revision{Overall, \sys achieves the best accuracy
and robustness across the selectivity spectrum.  In the tail, it outperforms {\sf
  MHIST} by $691\times$, {\sqlserv} by $114\times$, un-tuned (tuned) {\sf MSCN}
by $115\times$ ($33\times$), {\sf BayesNet} by $70\times$, {\sf Sample} by
$47\times$, and {\sf KDE-superv} by $21\times$}.
We next discuss takeaways from Table~\ref{table:standalone-combined}.

\textbf{Independence assumptions lead to orders of magnitude errors.} Estimators that assume full or partial independence between columns produce large errors, regardless of query selectivity or how good per-column estimates are.  These include {\sf Indep}, {\sf Postgres}, and {\sqlserv}, whose tail errors are in the $10^3-10^5 \times$ range. \sys's model is powerful enough to avoid this assumption, leading to better results.

\revision{\textbf{{\sf MHIST}} outperforms {\sf Indep} and {\sf Postgres} by over an order of magnitude.
However, its performance is limited by
the linear partitioning and uniform spread assumptions.
}

\revision{\textbf{{\sf BayesNet}} also does quite well, nearly matching
  supervised approaches, but is still significantly outperformed by \sys due to
  the former's uses of lossy discretization and conditional independence
  assumptions.
}


\textbf{Worst-case errors are much harder to be robust against.}
All estimators perform worse for low-selectivity queries or at worst-case
errors. For instance, in the high-selectivity regime {\sf Postgres}'s error is a reasonable $1.55\times$ at 95th, but becomes $1682\times$ worse at the maximum. Also, {\sf Sample} performs exceptionally well for high and medium selectivity queries, but drops off considerably for low selectivity queries where the sample has no hits.
{\sf MSCN} struggles since its supervised objective requires more training data to cover all possible low-selectivity queries.
\sys yields much lower (single-digit) errors at the tail, showing
the robustness that results from directly approximating the joint.

\textbf{KDE struggles with high-dimensional data.} {\sf KDE}'s errors are among the highest. The reason is that, the bandwidth vector found is highly sub-optimal despite tunings, due to (1) a large number of attributes in \dmv,
and (2) discrete columns fundamentally do not work well with the notion of ``distance'' in KDE~\cite{selectivity-kde}.  The method must rely on query feedback ({\sf KDE-superv}) to find a good bandwidth.


\textbf{MSCN heavily relies on its materialized samples for accurate prediction.}
Across the spectrum, its accuracy closely approximates {\sf Sample}.
{\sf MSCN-10K} has $3\times$ better tail accuracy than {\sf MSCN-base} due to access to 10$\times$ more samples, despite having the same network architecture and trained on the same 100K queries.
Both variants' accuracies drop off considerably for low-selectivity queries,
since, when there are no hits in the materialized sample, the model relies
solely on the query features to make ``predictions''.  \textsf{MSCN-0} which
does not use materialized samples performs
much worse, obtaining a max error of $6145\times$.

\subsubsection{Results on \conviva-A}

Based on \dmv results, we keep only the promising baselines for this dataset. Table~\ref{table:standalone-combined-conviva} shows that \sys remains best-in-class for a dataset with substantially different columns and a much larger joint size.

For this dataset, most estimators produce larger errors. This is because \conviva-A has a much larger joint space. \revision{{\sqlserv}, {\sf MHIST}, {\sf BayesNet}, and {\sf KDE-superv} exhibit $5\times$, $14\times$, $1.8\times$, and $2.6\times$ worse max error than before respectively.} 
{\sf MSCN-base}'s max error in the medium-selectivity regime is also $14\times$ worse.
As a non-parametric method covering the full joint space, {\sf Sample} remains a
robust choice.\removedfromrevision{ as the column count is scaled up.}

For \sys, since the sampler needs to cross more domains and a much larger joint
space, \sys-1000 becomes insufficient to provide single-digit error in all
cases.  However, a modest scaling of the number of samples to 4K decreases the
worst-case error back to single-digit levels.  This suggests that the
approximated joint is sufficiently accurate, and that the key challenge lies in
{\em extracting its information}.\\




\subsection{Robustness to Out-of-Distribution Queries}
\label{subsec:ood}
Our experiments thus far have drawn the filter literals (query centers) from the data.
However, a strong estimator must be robust to out-of-distribution (OOD)
queries where the literals are drawn from the entire joint domain,
which often result in no matching tuples. 
Table~\ref{table:ood} shows results on select estimators on 2K OOD queries on \dmv, where
$98\%$ have a true cardinality of zero. The supervised {\sf MSCN-10K}
suffers greatly (e.g., median is now $23\times$, up from the $1.51\times$ in Table~\ref{table:standalone-combined}) because it was trained on a set of in-distribution queries; at test time, out-of-distribution queries confuse the net.
{\sf KDE-superv}, a sampling-based approach, finds no hits in its sampled tuples, and therefore appropriately assigns zero density mass for all queries.
\begin{table}[htp]\centering \small%
\caption{{Robustness to OOD queries. Errors from 2,000 queries.}}
\begin{tabular}{@{} r l l l l @{}} \toprule
 {\textsc{Estimator}} & Median & 95th & 99th & Max \\ \midrule
 {\sf MSCN-10K} & 23 & 96 & 151 & 417 \\
{\sf KDE-superv} & 1.00 & 1.00 & 3.67 & 163 \\
 {\sf Sample} & 1.00 & 1.00 & 2.00 & 116 \\
 {\sys-2000} & 1.00 & 1.00 & 1.26 & 4.00 \\
\bottomrule
\end{tabular}
\label{table:ood}
\end{table}

Since \sys approximates the data distribution, it correctly learns that out-of-distribution regions have little or no density mass, outperforming KDE by $40\times$ and MSCN by $104\times$.

\subsection{Training Time vs. Quality}\label{subsec:training}

Compared to supervised learning, \sys is efficient to train: no past queries are required; we only need access to a uniform random stream of tuples from the relation. We also find that, surprisingly, it only takes a few epochs of training to obtain a sufficiently powerful \sys estimator.

Figure~\ref{fig:dmv-train-time-vs-divergence} shows how two quality
metrics, entropy gap and estimation error, change as training
progresses.  The metrics are calculated after each epoch (one pass over the data) finishes.  An epoch takes about 75 seconds and 50 seconds for \dmv and \conviva-A, respectively.
The number of progressive samples is set to 2K for \dmv and 8K for \conviva-A.

Observe that \sys quickly converges to a high goodness-of-fit both in terms of entropy gap and estimation quality.  For \dmv where a larger \sys model is used, 1 epoch of training suffices to produce the best estimation accuracy compared to all baselines (Table~\ref{table:standalone-combined}, last column).  For \conviva-A, 2 epochs yields the best-in-class quality and about 15 epochs yields the quality of single-digit max error.


\subsection{Estimation Latency}\label{subsec:latency}
Figure~\ref{fig:latency} shows \sys's estimation latency against other
baselines.  On both datasets \sys can finish estimation in around 5-10$\si{\ms}$ on
a GPU, which is faster than scanning samples (\textsf{Sample} and \textsf{MSCN})
and is competitive
with {\sqlserv}. We note the caveat that latencies for \pg and \sqlserv include
producing an entire plan for each query.

\revision{Naive progressive sampling requires as many model forward passes as the number
of attributes in the relation. With
  \sys's wildcard-skipping optimization (\secref{sec:wildcard-skipping}), however, we can skip the forward passes
  that generate distributions for the wildcard columns.  Hence, the number of
  forward passes required is the number of non-wildcard columns in each
  query. This effect manifests in the slightly slanted nature of \sys's CDF curves---queries that
  only touch a few columns are faster to estimate than those with a larger
  number of columns.  Latency tail is also well-behaved: on \dmv,
\sys-1000's median is at ${6.4\si{\ms}}$ vs. max at ${9.4\si{\ms}}$; on
\conviva-A, \sys-2000's median is at ${5.0\si{\ms}}$ vs. max at $9.7\si{\ms}$.}

\sys's estimation latency can be further minimized by engineering. \sys's
sampler is written in Python code and a general-purpose deep learning framework
(PyTorch); the resultant control logic overhead from interpretation can be removed by
using hand-optimized native code. Orthogonal techniques such as
half-precision, i.e., 32-bit floats quantized into 16-bit floats, would shrink
\sys's compute cost by half.

\pagebreak

\begin{figure}[htp!]
\centering
     \centering
     \begin{subfigure}[b]{0.49\columnwidth}
         \centering
\includegraphics[width=\columnwidth]{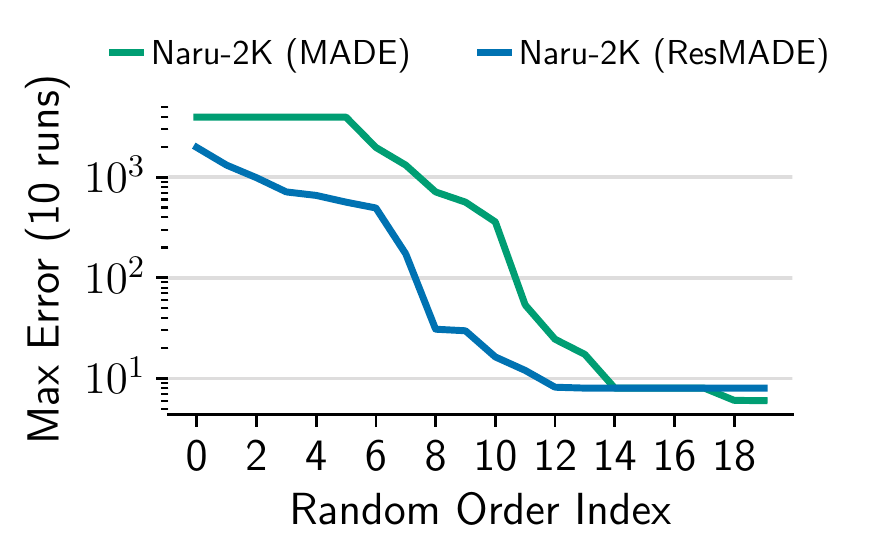}
\vspace{-.20in}
         \caption{\small{\dmv}}
     \end{subfigure}
     \hfill
     \begin{subfigure}[b]{0.49\columnwidth}
         \centering
\includegraphics[width=\columnwidth]{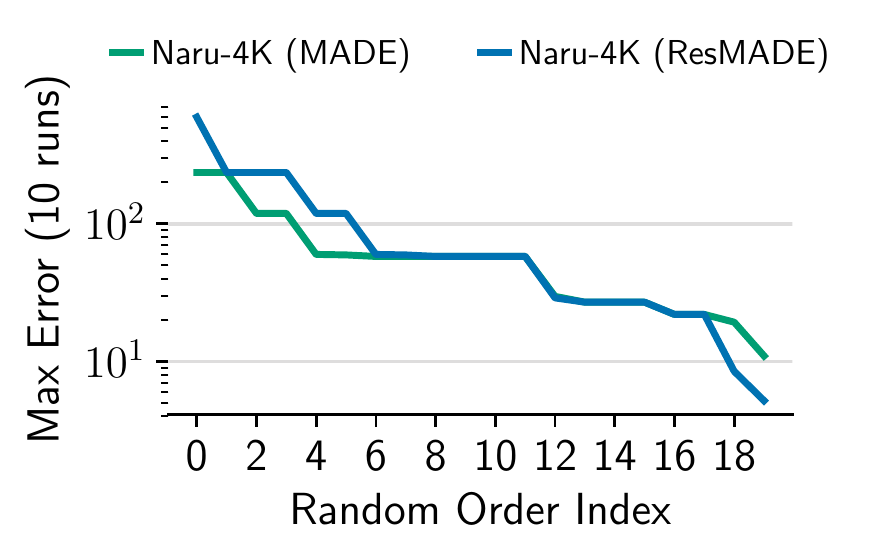}
\vspace{-.20in}
         \caption{\small{\conviva-A}}
     \end{subfigure}
\vspace{.00001in}
\caption{{\revision{{Variance of random orders.}
    Each dataset's 2000-query workload is run 10 times; the maximum of these 10
    max errors are shown. Order indices sorted by descending max error.}}}
\label{fig:rand-orders-eval}
\end{figure}

\begin{table}[hpt]
  \caption{\small{Larger model sizes yield lower entropy gap.
      Here we only consider scaling the hidden units of a MADE model.
    }\label{table:modelsize}}
  \centering \small
\begin{tabular}{@{}rllll@{}}
\toprule
\textsc{Architecture} & \textsc{Size (MB)} & Entropy gap, 5 epochs\\ \midrule
$32\times32\times32\times32$ & 0.6    & 4.23 bits per tuple \\
$64\times64\times64\times64$ & 1.1    & 2.25 bits per tuple \\
$128\times128\times128\times128$ & 2.7    & 1.01 bits per tuple \\
$256\times256\times256\times256$ & 3.8    & 0.84 bits per tuple \\

  \bottomrule
\end{tabular}

\end{table}

%
%
%


\begin{table}[htp]
  \centering \small
\caption{{{Comparison of autoregressive building blocks} (\secref{subsec:model-choice}).  Max error
      calculated by running \dmv's 2000-query workload 10 times (\sys-2000).
      FLOPs is the number of floating point operations required per
      forward pass per input tuple.}\label{table:ar-choice}}
\begin{tabular}{@{}rlllll@{}}
\toprule
          & {Params} & {FLOPs}  & \textsc{Ent. Gap}   & \textsc{Max Error}   \\ \midrule

  MADE       & 3.3M & {6.7M} & 0.59 & 8.0$\times$                  \\

  ResMADE & 3.1M & {6.2M} & 0.56 & 8.0$\times$ \\

  Transformer & 2.8M    & 35.5M  & 0.54 & 8.2$\times$ \\
  \bottomrule
\end{tabular}

\end{table}

\subsection{Variance Analysis}\label{eval:variance-reduction}
\begin{figure}[tp]
\centering
     \centering
     \begin{subfigure}[b]{0.98\columnwidth}
         \centering
\includegraphics[width=\columnwidth]{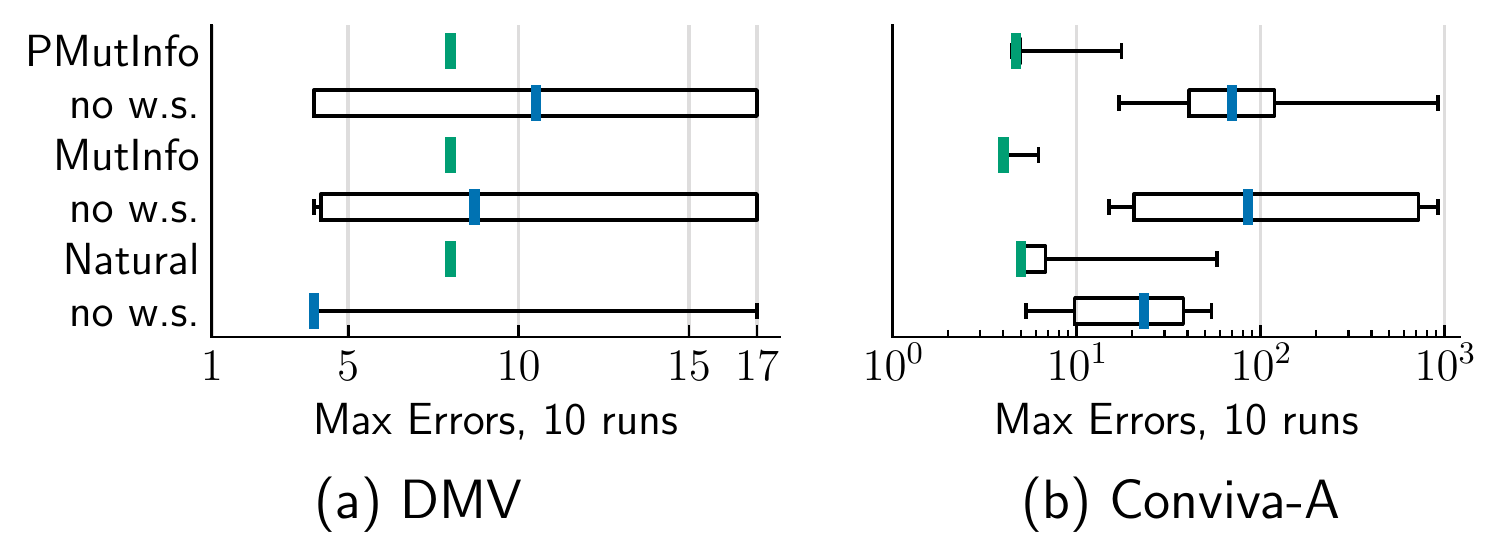}
     \end{subfigure}

\caption{Wildcard-skipping and heuristic orders. These orders have much lower variance than random orders.
      Ablation of wildcard-skipping is shown.
  Distributions of 10 max errors are plotted;
  whiskers denote min/max and bold bars denote medians.}
\label{fig:heuristic-orders-eval}
\end{figure}

\revision{

  \noindent {\bf Effect of random orders.}  Figure~\ref{fig:rand-orders-eval}
  shows the estimation variance of randomly sampled column orders.  We
  sample 20 random orders and train a \sys model on each, varying the autoregressive building block.
  The result shows that the choice of ordering
  does affect estimation variance in the extreme tail.  However, we found that
  on 99\%-tile or below, almost all orderings can reach
  single-digit errors. \\

  \noindent {\bf Effect of wildcard-skipping (\secref{sec:wildcard-skipping})
    and heuristic orders (\secref{sec:column-ordering}).}
  Figure~\ref{fig:heuristic-orders-eval} shows that
 the information-theoretic orders have much lower variance than randomly
 sampled ones.
The left-to-right order (\textsf{Natural}) is also shown
  for comparison.
  We also find that wildcard-skipping is critical in reducing max error variance
  by up to
  several orders of magnitude (e.g., {\sf MutInfo}'s max drops from $10^3$ to $<10$).

}

\subsection{Autoregressive Model Choice and Sizing}
\label{subsec:space}

In Table \ref{table:modelsize}, we measure the relationship between model size and entropy gap on \conviva-A. While larger model sizes yield lower entropy gaps, Figure \ref{fig:dmv-train-time-vs-divergence} shows that this can yield diminishing returns in terms of accuracy.  

Table~\ref{table:ar-choice} compares accuracy and (storage and computation) cost
of three similarly sized autoregressive building blocks. The results suggest that ResMADE and
regular MADE are preferable due to their efficiency.
We expect the Transformer---a more advanced architecture---to excel on datasets of larger scale.

\subsection{Understanding Estimation Performance}
\label{subsec:microbenchmarks}

\sys's accuracy depends critically on two factors: (1) the accuracy of the
density model; and (2) the effectiveness of progressive sampling.
This section seeks to understand the interplay between the two components and how each contributes to estimation errors. We do this by running microbenchmarks against the \conviva-B dataset, which has only 10K rows but has 100 columns for a total joint space of $10^{190}$. The small size of the dataset makes it possible to run queries against an emulated \textit{oracle model} with perfect accuracy by scanning the data. This allows us to isolate errors introduced by density estimation vs. progressive sampling.

\subsubsection{Robustness to Increasing Model Entropy Gap}
One natural question is: how accurate does the density model have to be?
One metric of modeling accuracy is \textit{the fraction of total probability mass assigned to observed data tuples}.
For example, a randomly initialized model will assign equal probability mass to all points in the joint space of tuples.
As training proceeds, it learns to assign higher probability to tuples actually present in the relation.
Under the simplifying assumption that all relation tuples are unique (as they are in \conviva-B), we can quantify this fraction as follows.
Suppose the model has an entropy gap of 2 bits; then, the fraction of probability mass assigned to the relation, $f$, satisfies $-\log_2 f = 2$, which leads to $f=25\%$. 

\begin{figure}[tp]
\centering
\includegraphics[width=\columnwidth]{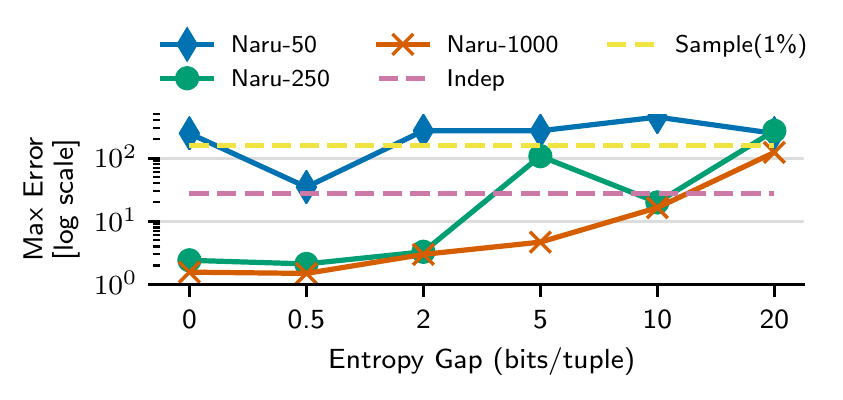}
\caption{\small{Accuracy of \sys as an artificial entropy gap is added to an oracle model for \conviva-B projected to the first 15 columns.
50 queries are drawn from the same distribution as in the macrobenchmarks.
\sys has the best accuracy for an entropy gap of less than 2 bits, though remains competitive up to a surprisingly large gap of 10 bits.
Variance of progressive sampling decreases dramatically when moving from 50 to 250 to 1000 samples.}}
\label{fig:bitgap}
\end{figure}

Figure \ref{fig:bitgap} shows that \sys achieves the best performance with a model entropy gap of 0-2 bits. A gap of lower than 0.5 bits does not substantially improve performance. This means that for the best accuracy, the model must assign between $25-100\percent$ of the probability mass to the empirical data distribution.
Surprisingly, \sys still outperforms baselines with up to 10 bits of entropy gap, which corresponds to less than $\approx 0.1\percent$ probability mass assigned.
We hypothesize that the range queries make such
modeling errors less critical,
because density errors of individual tuples could even out when estimating the density of the region as a whole.




\subsubsection{Robustness to Increasing Column Counts}

While the datasets tested in macrobenchmarks have a good number of columns,
using \conviva-B we test how well progressive sampling scales to 10$\times$ as
many dimensions. Figure~\ref{fig:manycols} shows that while the number of
columns does significantly increase the variance of estimates, the number of
progressive samples required to mitigate this variance remains tractable. A
choice of 1000 sample paths produces reasonable worst-case accuracies for up to
100 columns, and 10000 sample paths improves on that by a modest factor.

\subsubsection{Robustness to Data Shifts}
\label{subsec:refresh}

Lastly, we study how \sys reacts to data shifts.
We partition \dmv by a date column
into 5 parts.
We then ingest each partition in order, emulating the common practice of ``1 new partition per
day''.  Each estimator is built after seeing the first partition.
After a new ingest, we test the previously built estimators on
queries that touch all data ingested so far. The same query generator as
macrobenchmarks is used where the filters are drawn from tuples in the
first partition (true selectivities computed on all data ingested so far).
\begin{table}[hpt]
  \centering \small
\caption{\small{Robustness to data shifts. Errors from 200 queries.}}
\begin{tabular}{@{} r l l l l l @{}} \toprule
 {\textsc{Partitions Ingested}} & 1 & 2 & 3 & 4 & 5 \\ \midrule
 {\sys, refreshed}: max & 2.0 & 2.0 & 2.0 & 2.0 & 2.0 \\
  90\%-tile & 1.20 & 1.14 & 1.12 & 1.14 & 1.15 \\ \midrule
{\sys, stale}: max & 2.0 & 40.3 & 47.5 & 52.9 & 53.5 \\
  90\%-tile & 2.0 & 2.4 & 3.4 & 4.4 & 5.5 \\
\bottomrule
\end{tabular}\label{table:dataupdates}
\end{table}

Table~\ref{table:dataupdates} shows the results of (1) {\sys}, no model updates, (2) \sys, with
gradient updates on each new ingest. 
The model architecture is the same as in Table~\ref{table:standalone-combined};
8,000 progressive samples are used since we are interested in learning how much
imprecision or staleness presents in the model itself and not the effectiveness
of information extraction.
The results show that, \sys is able to handle queries on new data with reasonably good accuracy, even without having seen the new partitions.
The model has learned to capture the underlying data correlations so the degradation is graceful.


%% file: related.tex
\section{Related Work}

\sys builds upon decades of rich research on selectivity estimation and this section cannot replace comprehensive surveys~\cite{synopses_survey}.
Below, we highlight the most related areas.

 {\bf Joint approximation estimators.} Multidimensional histograms~\cite{poosala1997selectivity,gunopulos2005selectivity,muralikrishna1988equi,poosala_improved} can
 been seen as coarse approximations to the joint data distribution.
 Probabilistic relational models (PRMs)~\cite{getoor2001selectivity} rely on
  a Bayes Net (conditional independence DAG) to factor the joint into
  materialized conditional probability tables. Tzoumas \etal~\cite{tzoumas2011lightweight} propose a variant of
  PRMs optimized for practical use. Dependency-based
  histograms~\cite{deshpande2001independence} make partial or conditional
  independence assumptions to keep the approximated joint tractable (factors stored
  as histograms).
  \sys belongs to this family and applies recent advances from the deep unsupervised learning community.
  \sys does not make {\em any} independence assumptions; it directly models the joint distribution and lazily encodes all product-rule factors in a universal function approximator. 

  {\bf Query-driven estimators} are supervised methods that take advantage of past or training queries~\cite{chen_adaptive}.
  ISOMER~\cite{isomer} and STHoles~\cite{stholes} are two representatives that
  adopt feedback to improve histograms. LEO~\cite{leo} and CardLearner~\cite{cardlearner} use feedback to improve selectivity estimation of future queries.
  Heimel \etal~\cite{selectivity-kde} propose query-driven KDEs;
   Kiefer \etal~\cite{kiefer2017estimating} enhance them to handle joins.
  Supervised learning regressors~\cite{liu2015cardinality,kipf2018learned,dutt2019selectivity},
  some utilizing deep learning,
  have also been proposed. \sys, an unsupervised
  data-driven synopsis, is orthogonal to this family.  Our evaluation shows that
  full joint approximation yields accuracy much superior to two
  supervised methods.
  \removedfromrevision{However, the use of query feedback could serve as a
  beneficial fine-tuning mechanism.
}

  {\bf Machine learning in query optimizers.}  \sys can be used as a drop-in replacement
  of selectivity estimator used in ML-enhanced query optimizers.
Ortiz \etal~\cite{ortiz18_learn_state_repres_query_optim} learns query representation
to predict 
cardinalities, a {\em regression} rather
than our {\em generative} approach.  Neo~\cite{neo}, a {learned}
query optimizer, approaches cardinality estimation {\em indirectly}: embeddings for
all attribute values are first pre-trained; later, a network takes them as
input and
additionally learns to correct or ignore
signals from the embeddings.
This proposal, as well as reinforcement learning-based join optimizers (DQ~\cite{dq}, ReJOIN~\cite{rejoin}),
may benefit from \sys's improved estimates.

\begin{figure}[tp]
\centering
\includegraphics[width=\columnwidth]{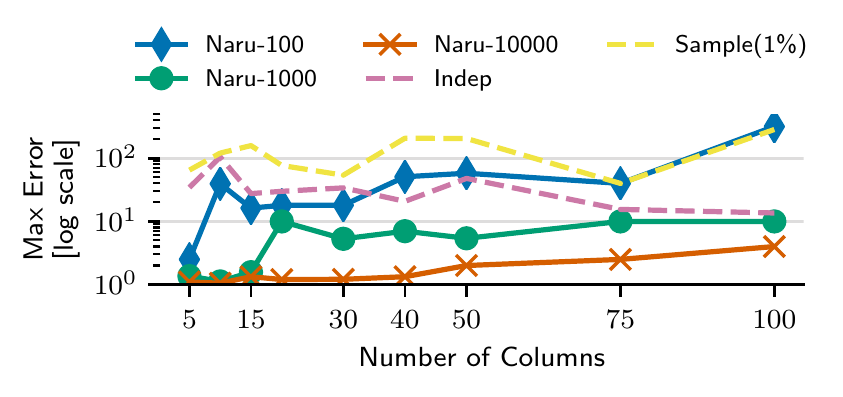}
\caption{\small{Accuracy of \sys as we add more columns from \conviva-B. We again use
  an oracle model (with 0 bits of entropy gap) and
  50 randomly generated queries. The number of predicates covers
  at most 12 columns. The number of progressive sample paths required to
  accurately query the model increases modestly with the number of columns, but
  remains tractable even as the joint data space reaches over
  $10^{190}$ (at 100 columns).}}
\label{fig:manycols}
\end{figure}


%% file: conclusion.tex
\section{Conclusion}
We have shown that
deep autoregressive models are highly accurate selectivity estimators.
They approximate the data distribution without any independence assumptions.
We develop a Monte Carlo integration scheme and associated variance reduction techniques that
efficiently handle challenging range queries. To the best of our knowledge,
these are
novel extensions to autoregressive models.
Our estimator, \sys,
exceeds the state-of-the-art in accuracy over several families of estimators.

\sys can be thought of as an unsupervised {\em neural synopsis}.
In contrast to supervised learning-based estimators, \sys enjoys drastically more efficient training
since there is no need to execute queries to collect feedback---it only
needs to read the data.
Learning directly from the underlying data
allows \sys to answer a much more general set of future queries and makes it
inherently robust to shifts in the query workload.  Our approach is
non-intrusive and can serve as an opt-in component inside an optimizer.

